\documentclass[fleqn,usenatbib]{mnras}

\usepackage{newtxtext,newtxmath}
\usepackage{pdflscape}
\usepackage[T1]{fontenc}

\DeclareRobustCommand{\VAN}[3]{#2}
\let\VANthebibliography\thebibliography
\def\thebibliography{\DeclareRobustCommand{\VAN}[3]{##3}\VANthebibliography}
\usepackage{graphicx}	% Including figure files
\usepackage{amsmath}	% Advanced maths commands
\usepackage{xfrac,nicefrac}

\newcommand{\Mdot}{\mbox{\,$\rm M_{\odot}$}}        % solar mass
\newcommand{\Zdot}{\mbox{\,$\rm Z_{\odot}$}}        % solar metallicity
\newcommand{\kms}{\,km s$^{-1}$}   			% kms-1
\newcommand{\aov}{$\alpha_{\rm ov}$}		% overshoot alpha
\newcommand{\fov}{$f_{\rm ov}$}		% overshoot alpha
\newcommand{\amlt}{$\alpha_{\rm mlt}$}		% mlt alpha
\title[WR yields]{New Wolf-Rayet wind yields and nucleosynthesis of Helium stars}
\author[E. R. Higgins et al.]{
Erin R. Higgins$^{1}$\thanks{E-mail: erin.higgins@armagh.ac.uk},
Jorick S. Vink$^{1}$,
Raphael Hirschi$^{2,3}$,
Alison M. Laird$^{4}$,
\newauthor
\& 
Andreas A.C. Sander$^{5}$
\\
$^{1}$Armagh Observatory and Planetarium, College Hill, Armagh BT61 9DG, N. Ireland\\
$^{2}$Astrophysics Group, Keele University, Keele, Staffordshire ST5 5BG, UK\\
$^{3}$Kavli Institute for the Physics and Mathematics of the Universe (WPI),\\ University of Tokyo, 5-1-5 Kashiwanoha, Kashiwa 277-8583, Japan\\
$^{4}$School of Physics, Engineering and Technology, University of York, York, YO10 5DD\\
$^{5}$Zentrum f{\"u}r Astronomie der Universit{\"a}t Heidelberg, Astronomisches \\Rechen-Institut, M{\"o}nchhofstr. 12-14, 69120 Heidelberg, Germany}

\date{Accepted 10 July 2024. Received 31 January 2024.}
\pubyear{2023}

\begin{document}

\maketitle
\begin{abstract}
\noindent
Strong metallicity-dependent winds dominate the evolution of core He-burning, classical Wolf-Rayet (cWR) stars, which eject both H and He-fusion products such as $^{14}$N, $^{12}$C, $^{16}$O, $^{19}$F, $^{22}$Ne and $^{23}$Na during their evolution. The chemical enrichment from cWRs can be significant. cWR stars are also key sources for neutron production relevant for the weak s-process. We calculate stellar models of cWRs at solar metallicity for a range of initial Helium star masses (12-50\Mdot), adopting the recent hydrodynamical wind rates from \cite{SV2020}. Stellar wind yields are provided for the entire post-main sequence evolution until core O-exhaustion. While literature has previously considered cWRs as a viable source of the radioisotope $^{26}$Al, we confirm that negligible $^{26}$Al is ejected by cWRs since it has decayed to $^{26}$Mg or proton-captured to $^{27}$Al. However, in \citet[][Paper I]{Higgins+23} we showed that very massive stars eject substantial quantities of $^{26}$Al, among other elements including N, Ne, and Na, already from the zero-age-main-sequence. Here, we examine the production of $^{19}$F and find that even with lower mass-loss rates than previous studies, our cWR models still eject substantial amounts of $^{19}$F. We provide central neutron densities (N$_{n}$) of a 30\Mdot\ cWR compared with a 32\Mdot\ post-VMS WR and confirm that during core He-burning, cWRs produce a significant number of neutrons for the weak s-process via the $^{22}$Ne($\alpha$,n)$^{25}$Mg reaction. Finally, we compare our cWR models with observed [Ne/He], [C/He] and [O/He] ratios of Galactic WC and WO stars.

\end{abstract}


\section{Introduction}

The chemical enrichment of galaxies relies on the nucleosynthesis and ejecta of stars which recycle material from their host environment and enrich their surroundings with fusion products either by stellar winds or supernovae. 
Characterized by their strong emission-line spectra, Wolf-Rayet (WR) stars \citep{WR67} are objects with particularly strong winds.
Many of the objects are core He-burning stars, nowadays called ``classical'' WR stars to distinguish them from other objects with the WR phenomenon \citep{Crow07}. Classical WR (cWR) stars are expected to form through a variety of channels due to mass loss and/or mixing, ranging from
% Classical Wolf-Rayet stars (cWR) are expected to form through
chemical mixing via rotation \citep{Yoon05, woosleyHeger06}, or large convective cores from VMS \citep[independent of rotation,][]{VinkHarries}; or via stripping, either self-stripping by main-sequence winds \citep{Conti80} or in binaries \citep{pacz67,Podsi92, Gilkis+2019, Klencki+2020, Laplace+2020, goetberg20}. Therefore, the subsequent high mass-loss rates of cWR stars have been predicted to be a large source of chemical feedback and enrichment in galaxies \citep[e.g.][]{meynetarnould00, binns05,mm12}. In particular, the radioisotope $^{26}$Al, which has been detected in the Galactic plane and is predicted to be crucial in the formation of our Solar System, has been attributed in some cases to the ejecta of cWR winds \citep{Arnould97,arnould06,gaidos09,tatischeff10,Fuji18}, while recent studies have shown alternative sources for $^{26}$Al \citep{limongichieffi06, brinkman19, martinet22, Higgins+23}. During core Helium (He) burning, cWRs efficiently fuse the H-processed $^{14}$N to the isotope $^{22}$Ne by double-$\alpha$ capture. The resulting $^{22}$Ne is an important source for the slow neutron-capture process (s-process) in massive stars. Indeed, the $^{22}$Ne($\alpha$,n)$^{25}$Mg reaction supplies a high neutron density for weak s-process reactions in post-H burning phases of evolution \cite{Frisch16, mm12}.

The mass-loss rates of cWR stars are critical in predicting accurate wind yields, and have developed significantly over the past decades. \cite{NugisLamers} provided an empirical mass-loss prescription based on the Galactic cWR population, suggesting that total $Z$, including $^{12}$C contributed to the driving of cWR winds. However, the self-enriched cWRs would therefore also maintain strong winds at lower $Z$ due to the $^{12}$C-production during core He-burning. \cite{Vink05} found that it was in fact the iron (Fe) abundance which was driving the winds of cWRs, meaning that lower $Z$ environments would eject less mass and collapse to form heavier black holes. This finding was important for the first gravitational-wave detections which measured black holes of $\sim$ 40\Mdot\ where the previous \cite{NugisLamers} would predict stellar black holes of 10-20\Mdot\ regardless of the host $Z$ environment. \cite{EV2006} explored the consequences of Z$_{\rm{Fe}}$-dependent cWR winds on the final masses, lifetimes and populations of cWRs, and is now implemented in some model grids \citep[e.g.][]{groh19}. More recently, \cite{SV2020} calculated hydrodynamically-consistent stellar atmospheres of cWRs further confirming the Fe-driving of cWR winds. In \cite{Higgins+21}, the implementation of this modern wind prescription led to the production of black hole progenitors with a wide mass range.

Observationally, WR stars are sorted in further subclasses based on prominent features in their (optical) spectrum. WN stars are characterized by prominent nitrogen lines and the absence of strong carbon lines. WC stars instead show prominent carbon emission lines while WO stars also show strong oxygen emission features. It has traditionally been predicted that the three subtypes also follow an evolutionary sequence \citep[WN-WC-WO; e.g.][]{maeder92}. However, since the core evolution cannot be directly inferred from the observed spectrum or abundances, the exact evolution status of each individual WN, WC, and WO star is difficult to constrain and remains unknown for the bulk of the population.

Beside He-burning cWR stars, the spectroscopic definition of a WN star can also be reached for H-burning stars which are massive and luminous enough to develop optically thick winds \citep{VG12}. At \Zdot, this applies to stars above $\sim$80-100\Mdot\ \citep{martins15,Sabh22} and these objects are called very massive stars \citep[VMS;][]{vinkbook}. Owing to their hydrogen, these stars are spectroscopically classified as WNh stars \citep{CrowWal11}. While this label is in principle also used for He-burning WN stars with remaining hydrogen, its usage without a specific subtype is often referring to VMS. At solar metallicity, the occurrence of hydrogen is further highly correlated with WN stars of a so-called ``late'' spectroscopic subtypes (WNL, meaning WN7 or later), while ``early'' (WNE, i.e., WN6 and earlier) stars are mostly hydrogen-free \citep[e.g.][]{Hamann06,Hamann+2019}. Therefore, the labels WNL and WNE have traditionally also been used to describe WN stars with and without hydrogen, but since this correlation disappears at subsolar metallicity, we refrain from using this convention.

In this work, we focus on hydrogen-free cWR stars, which encompasses the spectral types of H-free WNs, WCs and WOs. In the Milky Way, most of the 660 known WR stars \citep{rosslowecrow05} are cWRs. \cite{Hamann+2019} has provided stellar parameters of the single WN stars, with analysis of WC stars performed by \cite{Sander12}, and WO stars analysed by \cite{Tramper15} and later by \cite{aadland22}. The observed ratio of WC to WN stars has been of interest to the community due to the Z-scaling of this ratio which increases with host $Z$. \cite{NeugentMassey19} present an overview of the cWR populations in the Milky Way, M33, NGC6822, Large Magellanic Cloud (LMC) and Small Magellanic Cloud (SMC). \cite{Crow07} provides further details on the formation, evolution and populations of cWR stars. While spectroscopic analysis of cWR stars predominantly provides the surface He, C, N and O abundances, the forbidden Ne {\sc iv} lines can also estimate the surface neon (Ne) abundance. \cite{Dessart2000} provide estimates of Ne abundances for five WC stars in the Milky Way. 

In this work we present cWR, helium star models (Sect. \ref{method}) and provide stellar wind yields with a discussion of the relevant nucleosynthesis in Sect. \ref{yields}. We also include analysis of the central neutron production relevant for the weak s-process in Sect. \ref{sprocess}. A comparison between cWR stars and post-VMS Helium stars \citep[from Paper I, ][]{Higgins+23} is provided in Sect. \ref{vms}. Finally, we test the nucleosynthesis and resulting surface abundances of our cWR models against Galactic observations in Sect. \ref{obs} before presenting our conclusions in Sect. \ref{conclusions}.

\section{Method}\label{method}
In this work, we explore the evolution of Helium stars which have been completely stripped off their outer hydrogen envelope. Initially resembling surface abundances similar to observed, hydrogen-free WN stars, Helium star models are a frequently employed tool \citep[e.g.][]{pols02,mcclellannd16,Woosley19} to explore the evolution and impact of stars that lost their hydrogen envelope prior or close to the onset of central He burning. Therefore, Helium star models have been calculated using the one-dimensional stellar evolution code \texttt{MESA} \citep[v10398;][]{Pax11,Pax13,Pax15,Pax18,Pax19} for a grid of initial masses of 12\Mdot, 15\Mdot, 20\Mdot, 25\Mdot, 30\Mdot, 35\Mdot, 40\Mdot, 45\Mdot, and 50\Mdot. All calculations begin with a pre-He main sequence (MS), described in Sect. \ref{H_burn}, and evolve from the He-ZAMS until core O-exhaustion ($^{16}\rm{O}_{\rm{c}}$ $<$ 0.00001). We implement a nuclear reaction network which includes the relevant isotopes for evolution until the end of core O-burning. This nuclear network comprises the following 92 isotopes: n, $^{1, 2}$H, $^{3, 4}$He, $^{6, 7}$Li, $^{7, 9, 10}$Be, $^{8, 10, 11}$B, $^{12, 13}$C, $^{13-16}$N, $^{14-19}$O, $^{17-20}$F, $^{18-23}$Ne, $^{21-24}$Na, $^{23-27}$Mg, $^{25-28}$Al, $^{27-33}$Si, $^{30-34}$P, $^{31-37}$S, $^{35-38}$Cl, $^{35-41}$Ar, $^{39-44}$K, and $^{39-44,46, 48}$Ca. Our stellar models are computed with solar metallicity, where $X$ $=$ 0.720, $Y$ $=$ 0.266, and \Zdot\ $=$ 0.014 where the relative composition is adopted from \cite{asplund09}, provided in Table \ref{tab:abundances}. We avail of the OPAL opacity tables from \cite{RogersNayfonov02}, and adopt nuclear reaction rates from the JINA Reaclib Database \citep{Cyburt10}. 

\begin{table}
    \centering
    \caption{Initial abundances of chemical elements in mass fractions for our grid of models at \Zdot.}
    \begin{tabular}{c|c|c|c}
    \hline
      Isotope &  Mass fraction & Isotope & Mass fraction\\
      \hline \hline
      $^{1}$H   & 0.719986 & $^{20}$Ne & 1.356E-3\\
       $^{2}$H & 1.440E-5 & $^{22}$Ne & 1.097E-4 \\
        $^{3}$He  & 4.416E-5 & $^{23}$Na & 2.9095E-5\\
         $^{4}$He & 0.266 & $^{24}$Mg & 4.363E-4\\
          $^{12}$C & 2.380E-3 &$^{25}$Mg & 5.756E-5\\
          $^{14}$N & 7.029 E-4 & $^{26}$Mg & 6.585E-5\\
        $^{16}$O & 6.535E-3 & $^{27}$Al & 5.051E-5 \\
        $^{18}$O & 1.475E-5 & $^{28}$Si & 5.675E-4\\
        $^{19}$F & 3.475E-7 & $^{32}$S & 2.917E-4\\
        \hline
    \end{tabular}

    \label{tab:abundances}
\end{table}

The mixing-length-theory (MLT) of convection describes the treatment of convection in our models, where we apply an efficiency of \amlt $=$ 1.67 \citep{arnett19}. The Schwarzschild criterion defines the convective boundaries in our models and as such we do not implement semiconvective mixing. For convective boundary mixing (CBM), we include the exponential decaying diffusive model of \citet{Freytag1996} \citep[see also][]{Herwig2000} with \fov $=$ 0.03 (corresponding to \aov $\simeq$ 0.3) for the top of convective cores and shells, and with \fov $=$ 0.006 for the bottom of convective shells. In order to evolve these models to late evolutionary stages, we apply convection in superadiabatic layers via the \texttt{MLT++} prescription which aids numerical convergence. The temporal resolution of our models has been set with \texttt{varcontrol}\texttt{target} $=$ 0.0001, and a corresponding spatial resolution of \texttt{mesh}\texttt{delta} $=$ 0.5.

During core He, C and O-burning phases of each model we adopt the physically motivated mass-loss rates based on hydrodynamically consistent stellar atmospheres from \cite{SV2020}. As previously implemented in \cite{Higgins+21}, we adopt the following $\dot{M}(L)$-recipe 
\begin{equation}
    \label{Seq}
    \dot{M}_{\mathrm{SV20}} = \dot{M}_{10}\ \left(\log \frac{L}{L_{0}}\right)^\alpha \left(\frac{L}{10 \cdot L_{0}}\right)^{3/4} 
\end{equation} 
provided by \cite{SV2020}, with coefficients $\dot{M}_{10}$ $=$ -4.075, $L_{0}$ $=$ 5.043 and $\alpha$ $=$ 1.301. While additions have been provided by \cite{Sander23} on the T-dependency of mass-loss rates, we find our stellar models to be within the appropriate T range where the prior rates from \cite{SV2020} are applicable. While mass-loss rates beyond core He-burning are still uncertain, and as the post-He timescales are only $\sim$1.5\% of core He-burning, the overall wind yields should not be overly impacted as long as late-stage mass loss does not scale completely different from what we assume.For sufficient wind mass-loss, the surface abundances will change from a WN-like composition to one that resembles WC or WO stars. Since we do not adopt different mass-loss recipes for these regimes, we do not need any abundance criteria in our evolutionary models and only define them for the purpose of comparing with observations in Sect. \ref{obs}.

\subsection{Towards pure Helium star evolution}\label{H_burn}
To calculate our grid of He star models, we evolve H-ZAMS models towards the He-ZAMS via extreme mixing, which promotes bluewards evolution by dredging additional H  into the core. Rather than inducing rapid rotation, we employ an artificially large increase of the convective core by exponential overshooting. We include core convective overshooting above the H-burning core with a diffusive exponential method for values of $f_{\rm{ov}}$ up to 0.9. In Nature, pure Helium stars could be achieved through various paths, including strong winds, rapid rotation, and/or binary evolution. Rotation is included in all models during core H-burning with angular momentum transport and chemical mixing coefficients from \cite{Heger00}, with an initial rotation rate set to 20\% critical at the H-ZAMS. While increased mixing by rotation promotes evolution towards the He-ZAMS, the core He-burning models have sufficiently spun down in the first $\sim$ 10,000 years due to angular momentum loss by stellar winds such that the rotation rates are all reduced to $\leq$ 150\kms \citep{vink11b,GVHL12}.
\\
% \textbf{Rationale : Ref 3}\\
We implement zero mass loss during core H-burning in order to create pure He star models which remain massive enough on the He-ZAMS to probe the range of masses 12-50\Mdot. Crucially, by evolving from the H-ZAMS rather than forming a pure Helium star on the He-ZAMS, we follow the nucleosynthesis from H-burning such that the production of isotopes (e.g. $^{4}$He, $^{14}$N, $^{26}$Al) are modelled explicitly. This method allows for accurate mapping of Helium star yields, where the star has been stripped and begins core He-burning as a pure Helium star, without prior impositions of how the cWR star became stripped \citep[see also][]{joris24}. We note that while the yields of some isotopes may be affected by mass loss on the MS (e.g. $^{14}$N), we consider here the reprocessing of such H-products during the core He-burning phase (e.g. into $^{12}$C or $^{22}$Ne). The ejected masses, yields, and nucleosynthesis detailed in this paper are relevant for single and binary star models which may be implemented in population synthesis or galactic chemical evolution (GCE) models. While in some scenarios the effects of stripping towards forming a pure Helium star may occur after core He-burning has initiated, we do not explore the cases which involve partial stripping or envelope stripping at various stages during core He-burning, but focus on the pure Helium star case. With our modelling approach, we implicitly assume that cWR stars have lost all of their hydrogen envelope. While there are observed cWR stars with remaining hydrogen, the bulk of the observed cWR population at \Zdot\ is clearly identified as He-burning and fulfils this criterion \citep[e.g.][]{Hamann+2019}, in contrast to lower metallicity environments \citep[e.g.][]{Hainich14,Hainich15}. We thus do not cover WN stars with considerable surface H.

\begin{figure*}
    \centering
    \includegraphics[width = \textwidth]{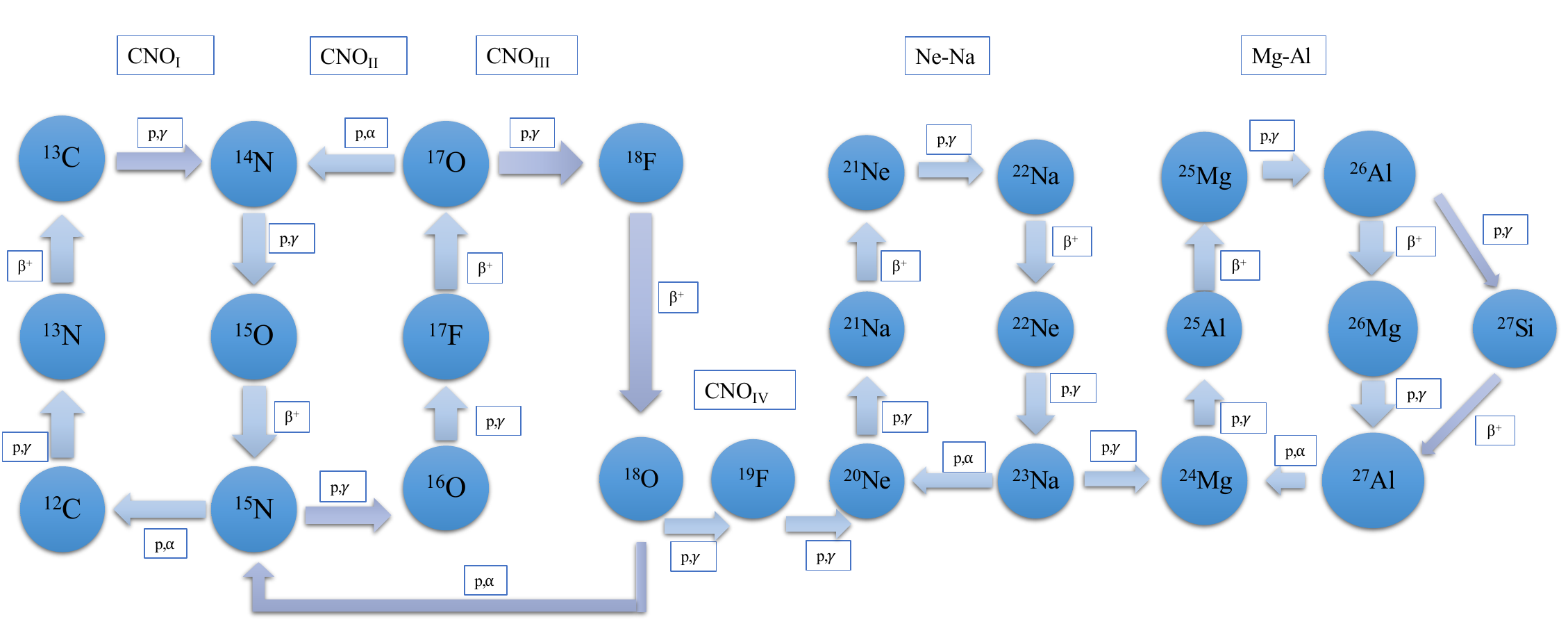}
    \caption{Illustrative flow diagram of the key isotopes and reaction flows of the CNO (I-IV), Ne-Na and Mg-Al cycles, during core H-burning.}
    \label{cno}
\end{figure*}
\begin{figure*}
    \centering
    \includegraphics[width = \textwidth]{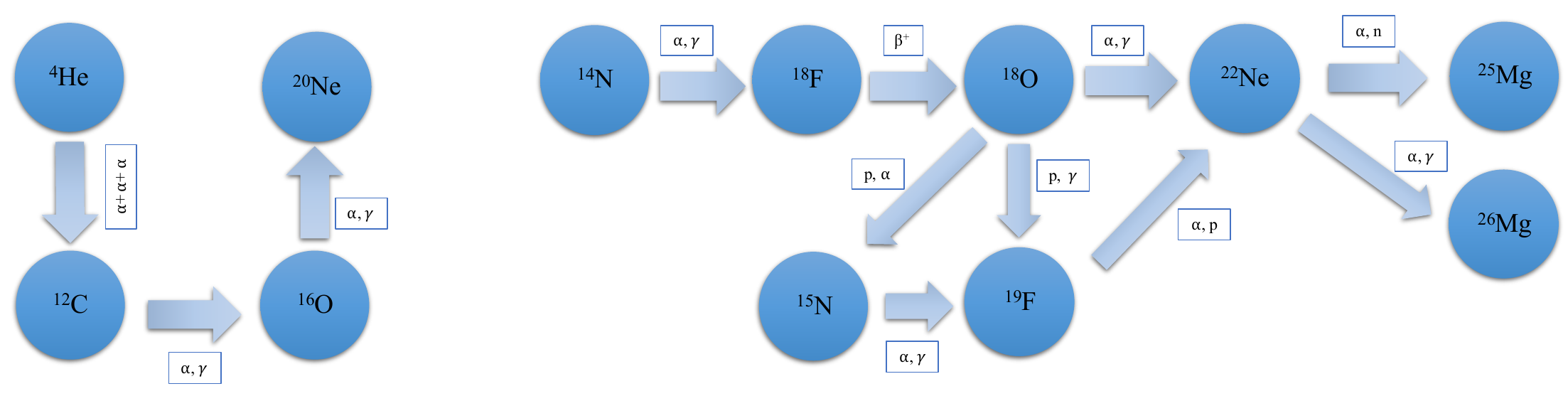}
    \caption{Diagram of the key isotopes and reaction flows of $\alpha$-capture reactions during core He-burning.}
    \label{alphacap}
\end{figure*}

\begin{table*}
    \centering
    \begin{tabular}{ccccccccccc}
    \hline
         M$_{\rm{He-ZAMS}}$&  M$_{\rm{He-TAMS}}$&  M$_{\rm{CO}}$&  M$_{\rm{f}}$& $\tau$$_{\rm{He}}$ & $\tau$$_{\rm{post-He}}$ & Tc$_{\rm{He-ZAMS}}$&  Tc$_{\rm{He-HAMS}}$&  Tc$_{\rm{He-TAMS}}$&  Tc$_{\rm{C-TAMS}}$& Tc$_{\rm{O-TAMS}}$\\
\hline \hline
         12&  11.715&  9.136&  11.684&    0.531	&8352.37  &0.076	&0.204	&0.314	&1.136	&2.654\\
         15&  13.322&  10.627&  13.268&   0.467	&7143.76  &0.078	&0.207	&0.317	&1.162	&1.442\\
         20&  15.408&  12.554&  15.319&   0.418	&6358.12  &0.081	&0.211	&0.322	&1.195	&2.634\\
         25&  17.239&  14.240&  17.119&   0.390	&5948.49  &0.082	&0.214	&0.325	&1.221	&2.610\\
         30&  18.918&  15.813&  18.771&   0.371	&5610.17  &0.084	&0.217	&0.328	&1.238	&2.680\\
         35&  20.502&  17.287&  20.328&   0.358	&5360.80  &0.085	&0.219	&0.331	&1.252	&2.231\\
         40&  22.012&  18.708&  21.813&   0.347	&5184.21  &0.086	&0.220	&0.333	&1.265	&2.832\\
         45&  23.464&  20.070&  23.240&   0.339	&5024.13  &0.086	&0.222	&0.334	&1.276	&2.879\\
         50&  24.871&  21.386&  24.623&   0.332	&4881.42  &0.087	&0.223	&0.336	&1.286	&2.897\\
         \hline
    \end{tabular}
    \caption{Stellar parameters for the model grid with initial masses ranging from 12-50\Mdot. The total masses at the end of core He-burning (M$_{\rm{He-TAMS}}$), CO core masses at the end of core He-burning (M$_{\rm{CO}}$), and final masses (M$_{\rm{f}}$) are provided. The burning timescales are provided for core He-burning ($\tau$$_{\rm{He}}$) in Myrs, and post He-burning (core C-burning and O-burning, $\tau$$_{\rm{post-He}}$) in years. Similarly, the core temperatures (in GK) are provided for the He-ZAMS (Tc$_{\rm{He-ZAMS}}$), mid-He-burning (Tc$_{\rm{He-HAMS}}$), end of core He-burning (Tc$_{\rm{He-TAMS}}$), C-burning (Tc$_{\rm{C-TAMS}}$) and O-burning phases (Tc$_{\rm{O-TAMS}}$).}
    \label{tab:properties}
\end{table*}

Table \ref{tab:properties} details the stellar masses at the end of core He-burning and the end of core O-burning, while also providing the M$_{\rm{CO}}$ core mass at the end of core He-burning. The final masses of our model grid range from 9-21\Mdot\ with carbon-oxygen (CO) cores which are $\sim$80\% of the total mass of these stripped star models. The timescales of core He-burning and post He-burning phases (C and O) are included, alongside the central temperatures at the start, middle and end of core He-burning, as well as at the end of core C and O burning. The central temperatures are systematically higher at each stage for increasing stellar mass leading to more efficient nuclear burning. For all masses, the core C-burning timescale is $\sim$ 1.5\% of that of the core He-burning phase. We illustrate the evolution of our model grid in a Hertzsprung-Russell diagram in Fig. \ref{fig:HRD}, and show the mass evolution of our grid in Fig. \ref{fig:Mt} for reference.

% However, to provide a complete picture of the nucleosynthesis in Helium stars, we have provided comprehensive tables of ejected masses in Tables \ref{tab:92yield30} and \ref{tab:92yield200}, which are discussed in Appendix \ref{92yieldapp}. These tables have been produced for the 30\Mdot\ Helium star model, and for a M$_{\rm{H-ZAMS}}$ $=$ 200\Mdot\ model (with a comparable post-VMS 32\Mdot\ Helium star) from Paper I.
% Figure \ref{fig:neut_Xc} 

\section{Nucleosynthesis and wind yields}\label{yields}
We calculate net wind yields and ejected masses for our grid of cWR models. While chemical yields are a key input for GCE models, the ejected masses provide crucial information about how stars enrich their host environment with solar masses of nucleosynthesised material through strong winds. We adapt the relations from \citet{hirschi05, Higgins+23} for our yield calculations. The net wind yield calculated for a star of initial mass, $m$, and isotope, $i$, is:
\begin{equation}\label{yieldeq}
    m^{\rm wind}_{i} = \int_{0}^{\tau(m)} \dot{M}(m, t) \,[{X}^{S}_{i}(m, t) - {X}^0_{i}] \,dt 
\end{equation}
where $\dot{M}$ is the mass-loss rate, ${X}^{S}_{i}$ is the surface abundance of a given isotope, and ${X}^0_{i}$ is the initial abundance of a given isotope at the H-ZAMS. In this method, the correct feedback from the abundances at star formation is mapped accounting for the H-synthesised isotopes. The yields are then integrated from the beginning of core He-burning until $\tau(m)$, the end of core O-burning. 

We also calculate ejected masses, $EM$ of each isotope, $i$, by:
\begin{equation}\label{EMeq}
    EM_{im} = \int_{0}^{\tau(m)} \dot{M} \, {X}^{S}_{i}(m, t) \,dt .
\end{equation}

We present the complete table of ejected masses (top) and wind yields (bottom) in solar mass units for our model grid in Table \ref{tab:yields}. Given that our models have been calculated with a nuclear network of 92 isotopes, we focus on 14 key isotopes in Table \ref{tab:yields} for all models, and provide a table of ejected masses for 22 isotopes for a representative 30\Mdot\ model in Table \ref{tab:92yield30}. 

\subsection{Nucleosynthesis until core O-exhaustion}

During core H-burning, the CNO cycle leads to a pile up of $^{14}$N since the $^{14}$N(p,$\gamma$) reaction is the slowest reaction in the CNO-I cycle, and the CN-cycle (or CNO-I) is much faster than the CNO-II cycle. $^{15}$N is being destroyed and so decreases during core H-burning but $^{15}$N does start the second CNO cycle by producing $^{16}$O through proton-capture, allowing the $^{16}$O-reservoir to be available for the second CNO-cycle (producing more $^{14}$N and $^{4}$He). $^{15}$N increases at the end of core H-burning due to the CNO-III cycle via $^{18}$O(p, $\alpha$)$^{15}$N. This only occurs late in core H-burning since the CNO-III cycle is significantly slower than the CN or CNO-II cycles. We provide a schematic of the reaction flows through each of the CNO cycles in Fig. \ref{cno} for reference.

Secondary cycles also occur during H-burning which affect abundant isotopes of Ne, Na, Mg and Al, via the Ne-Na and Mg-Al cycles (see Fig. \ref{cno}). The Ne-Na cycle processes the initial $^{20}$Ne into $^{22}$Ne and $^{23}$Na before returning to $^{20}$Ne again. Therefore, the surface $^{20}$Ne abundance remains relatively constant throughout the evolution of cWR stars. Similarly, the Mg-Al cycle which occurs during core H-burning, converts $^{24}$Mg to $^{25}$Al - $^{25}$Mg - $^{26}$Al before decaying to $^{26}$Mg or proton-captures to $^{27}$Al via $^{27}$Si. 

% At the end of core H-burning, and the end of CNO-III, $^{18}$O proton-captures to $^{19}$F and then quickly at the ignition of He $\alpha$-captures to $^{22}$Ne. However, the much more abundant $^{14}$N also $\alpha$-captures leading to substantially abundant $^{22}$Ne, almost instantaneously. 

% During core H-burning, $^{19}$F decreases as it has proton-captured to $^{20}$Ne. 

Figure \ref{alphacap} illustrates the main $\alpha$-capture reactions which take place during core He-burning. At the onset of core He-burning, the H-processed $^{4}$He produces $^{12}$C through the triple-$\alpha$ reaction, before the increased C abundance and increased central temperature activate the $^{12}$C($\alpha$, $\gamma$)$^{16}$O reaction, where $^{16}$O($\alpha$, $\gamma$)$^{20}$Ne produces a modest amount of $^{20}$Ne. The resulting CO core at core He-exhaustion plays a key role in the compactness of the stellar core and explodability \citep{oconnor,Farmer+2019}. The abundant $^{14}$N present during core He-burning is synthesised to $^{18}$F which in turn transforms to $^{18}$O through $\beta$$^{+}$ decay, before $\alpha$-capturing to $^{22}$Ne, or proton-capturing to $^{19}$F. This abundant $^{22}$Ne leads to two competing reactions, the ($\alpha$, n)$^{25}$Mg which produces neutrons, and the ($\alpha$, $\gamma$)$^{26}$Mg reaction. The build-up of $^{15}$N from CNO-III via $^{18}$O(p, $\alpha$)$^{15}$N leads to $\alpha$-captures during core He-burning which results in a steep increase in $^{19}$F, which in turn $\alpha$-captures to produce $^{22}$Ne \citep[e.g.][]{arnett85,chieffi98}.

% During core C-burning $^{24}$Mg and $^{16}$O are produced via the $^{20}$Ne($\alpha$, $\gamma$)$^{24}$Mg and $^{20}$Ne($\gamma$, $\alpha$)$^{16}$O reactions (though these reactions are highly T-dependent), while also producing $^{16}$O, $^{20}$Ne and $^{23}$Na through $\alpha$ and proton captures. $^{24}$Mg and $^{16}$O are also produced subsequently during Ne-burning where photo-disintegration reactions occur. 

% {\it Can I suggest the following instead of the above paragraph:}\\
During core C-burning $^{20}$Ne and $^{23}$Na are produced via the $^{12}$C($^{12}$C,$\alpha$)$^{20}$Ne and 
$^{12}$C($^{12}$C,p)$^{23}$Na reactions \citep{thiele85, iliadis}. 
Subsequent proton and $\alpha$ capture reactions on $^{23}$Na and $^{16}$O also produce $^{20}$Ne. Additional proton captures also lead to $^{22}$Ne, $^{23}$Na,  
$^{24}$Mg, $^{26}$Al and $^{27}$Al.  
Once the $^{12}$C is exhausted, core Ne-burning is initiated by the photo-disintegration reaction $^{20}$Ne($\gamma$, $\alpha$)$^{16}$O. The resulting $\alpha$-particles are captured by $^{16}$O as well as by $^{20}$Ne, $^{23}$Na and $^{24}$Mg. Oxygen burning consists of a network of reactions, initiated by $^{16}$O $+$ $^{16}$O fusion. The resulting $^{32}$S is highly excited and many exit channels are open through the emission of light particles. The protons, neutrons and $\alpha$-particles released are quickly captured. The final composition at oxygen exhaustion is dominated by $^{28}$Si and $^{32}$S . \\

\begin{landscape}
\begin{table}
    \centering
    \begin{tabular}{c|c|c|c|c|c|c|c|c|c|c|c|c|c|c|c|c|c}
    \hline
        $M_{\rm{i}}/\rm{M}_{\odot}$ & $^{1}$H & $^{4}$He & $^{12}$C & $^{14}$N & $^{16}$O &$^{19}$F & $^{20}$Ne & $^{22}$Ne & $^{23}$Na &$^{25}$Mg &$^{26}$Mg &$^{26}$Al &$^{27}$Al &  $^{28}$Si \\
        \hline \hline
% 8&6.4E-8&7.3E-5&7.1E-9&6.1E-7&3.7E-9&1.2E-7&7.9E-10&1.8E-8&9.3E-9&1.1E-10&3.8E-9&4.3E-8\\
12&3.8E-5&3.1E-1&3.7E-5&2.6E-3&1.4E-5&   5.2E-12   &4.9E-4&3.3E-6&8.3E-5&   3.7E-8 &3.7E-5&1.3E-6&1.7E-5&1.8E-4\\
15&5.0E-5&1.7&2.2E-4&1.4E-2&7.6E-5&  2.9E-11  &2.7E-3&2.0E-5&4.7E-4&  2.6E-7  &1.9E-4&1.0E-5&1.1E-4&1.0E-3\\
20&7.1E-5&4.5&7.7E-2&3.4E-2&3.6E-3&  4.2E-6  &7.1E-3&7.2E-3&1.3E-3&  6.1E-6  &4.8E-4&4.6E-5&3.2E-4&2.7E-3\\
25&1.1E-4&6.7&9.1E-1&3.8E-2&1.4E-1&  2.4E-5  &1.2E-2&4.2E-2&2.2E-3&  1.1E-4  &9.6E-4&8.1E-5&5.6E-4&4.6E-3 \\  
30&1.3E-4&8.8&1.9&4.2E-2&4.0E-1&  4.2E-5  &1.7E-2&7.9E-2&3.2E-3& 3.9E-4   &1.7E-3&1.2E-4&8.2E-4&6.5E-3\\
\hline
32$^{\dagger}$&1.3E-1	&13.6	&2.9	&6.7E-2	&6.1E-1	&1.7E-5&	2.6E-2&	1.2E-1&	5.7E-3	&7.8E-4&	1.5E-3&	4.7E-4	&1.4E-3&	1.0E-2\\
\hline
35&1.5E-4&10.8&2.9&4.5E-2&7.5E-1&  5.9E-5  &2.2E-2&1.2E-1&4.3E-3&  8.8E-4  &2.7E-3&1.5E-4&1.1E-3&8.5E-3\\
40&1.8E-4&12.9&3.8&4.8E-2&1.2&  7.6E-5  &2.8E-2&1.6E-1&5.4E-3&  1.6E-3  &3.9E-3&1.8E-4&1.4E-3&1.1E-2\\
\hline
40$^{*}$&1.8E-4&	17.5&	6.1&	4.8E-2&	1.2&1.3E-4&	3.8E-2	&2.5E-1&	7.5E-3&1.3E-3&	4.5E-3	&1.8E-4&	1.9E-3&	1.5E-2\\
\hline
45&2.6E-4&15.0&4.8&5.1E-2&1.6&  9.4E-5  &3.4E-2&2.0E-1&6.5E-3& 2.5E-3   &5.5E-3&2.1E-4&1.7E-3&1.3E-2\\
50&2.1E-4&17.2&5.8&5.3E-2&2.0&  1.1E-4  &4.0E-2&2.4E-1&7.6E-3& 3.7E-3   &7.3E-3&2.3E-4&2.0E-3&1.5E-2\\
%       \hline
% 50*   & 5.698 & 2.105 & 0.014 & 4.847E-3 & 0.052 & 0.013 & 1.044E-3 & 2.251E-4 & 5.850E-4 & 0 & 3.033E-4 &  4.598E-3\\
% 100*   & 27.617 & 51.069 & 3.203 & 0.476 & 0.808 & 0.130 & 0.135 & 0.019 & 6.538E-3 & 2.678E-3 & 4.699E-3 & 0.049\\
      \hline
      \hline
% 8 &0&0&0&0&0&0&0&0&0&0&0&0\\
12&-2.27E-1&2.27E-1&-5.32E-4&2.41E-3&-2.04E-3& -1.55E-7   &-2.85E-5&-3.82E-5&7.40E-5&   -2.03E-5    &1.35E-5&1.25E-6&6.53E-6&2.37E-8\\
15&-1.25&1.25&-2.90E-3&1.32E-2&-1.12E-2&  -8.49E-7  &-1.70E-4&-2.08E-4&4.20E-4&   -1.11E-4   &6.32E-5&1.03E-5&4.39E-5&1.65E-7\\
20&-3.37&3.29&6.89E-2&3.11E-2&-2.70E-2& 1.93E-6   &-5.03E-4&6.61E-3&1.17E-3&    -2.96E-4   &1.34E-4&4.64E-5&1.43E-4&5.98E-7\\
25&-5.68&4.63&8.95E-1&3.35E-2&8.35E-2& 1.98E-5  & -8.88E-4&4.06E-2&2.01E-3&   -4.02E-4   &3.74E-4&8.08E-5&2.61E-4&-3.98E-8\\
30&-8.09&5.79&1.86&3.51E-2&3.27E-1&  3.63E-5  &-1.25E-3&7.70E-2&2.92E-3&  -3.39E-4  &8.49E-4&1.17E-4&3.91E-4&-4.45E-6\\
\hline
% 32 & -6.13E-1&	-2.97	&2.93&	-7.90E-2&	6.13E-1&	1.74E-5&	1.06E-4&	1.22E-1	&7.12E-5	&7.85E-4&	6.86E-4&	-5.15E-4&	2.76E-5&	-1.21E-5\\
32$^{\dagger}$ & -12.5&	8.96	&2.90&	5.63E-2&	4.99E-1&	8.78E-6	&-2.37E-3	&1.20E-1&	5.23E-3&	-3.51E-4&	2.18E-4&	4.67E-4&	7.11E-4&	-7.11E-6\\
\hline
35&-10.6&6.94&2.83&3.62E-2&6.50E-1&  5.20E-5 &-1.53E-3&1.15E-1&3.87E-3&  -7.15E-5   &1.58E-3&1.50E-4&5.29E-4&-1.40E-5\\
40&-13.1&8.09&3.81&3.68E-2&1.03&  6.74E-5  &-1.68E-3&1.54E-1&4.86E-3&   4.14E-4   &2.57E-3&1.80E-4&6.73E-4&-2.92E-5\\
\hline
40$^{*}$&-18.2&	10.8&6.1&3.29E-2&1.09&1.1E-4&-2.96E-3&	2.44E-1&6.07E-3&-3.0E-4&2.59E-3&	1.81E-4	&9.52E-4&	-1.24E-5\\
\hline
45&-15.7&9.24&4.77&3.72E-2&1.45& 8.32E-5  & -1.68E-3&1.93E-1&5.87E-3&   1.12E-3   &3.85E-3&2.07E-4&8.21E-4&-5.01E-5\\
50&-18.3&10.4&5.73&3.76E-2&1.89& 9.96E-5  & -1.52E-3&2.33E-1&6.91E-3&  2.02E-3    &5.41E-3&2.32E-4&9.73E-4&-7.66E-5\\

% 12&-2.9E-4&2.9E-4&4.7E-6&-7.0E-6&-1.7E-7&-5.0E-8&1.8E-7&-1.0E-7&-1.5E-7&-4.1E-8&1.9E-7&3.1E-10\\
% 15&-1.6E-3&1.6E-3&4.1E-5&-5.6E-5&-8.9E-7&-7.4E-8&9.3E-7&-8.4E-7&-1.1E-6&-2.3E-8&1.2E-6&2.2E-9\\
% 20&-5.0E-3&-7.8E-2&7.7E-2&-4.8E-3&3.4E-3&4.0E-8&7.2E-3&-7.9E-6&9.5E-6&-6.2E-6&1.8E-6&2.2E-8\\
% 25&-1.1E-2&-1.1&9.1E-1&-2.7E-2&1.4E-1&1.2E-5&4.2E-2&-2.8E-5&2.4E-4&-5.6E-5&-6.7E-6&-1.2E-6\\
% 30&-1.6E-2&-2.3&1.9&-5.1E-2&4.0E-1&9.0E-5&7.8E-2&-3.9E-5&7.5E-4&-1.4E-4&-2.2E-5&-6.4E-6\\
% 35&-2.3E-2&-3.6&2.9&-7.7E-2&7.5E-1&2.9E-4&1.2E-1&-4.4E-5&1.5E-3&-2.5E-4&-4.3E-5&-1.7E-5\\
% 40&-3.7E-2&-5.0&3.8&-1.0E-1&1.2&6.3E-4&1.6E-1&-4.8E-5&2.6E-3&-3.8E-4&-6.7E-5&-3.3E-5\\
% 45&-6.0E-2&-6.4&4.8&-1.3E-1&1.6&1.1E-3&2.0E-1&-5.5E-5&4.0E-3&-5.2E-4&-9.3E-5&-5.5E-5\\
% 50&-4.6E-2&-7.9&5.8&-1.6E-1&2.1&1.8E-3&2.4E-1&-4.5E-5&5.7E-3&-6.8E-4&-1.4E-4&-8.3E-5\\
%       \hline
% 50*   & -5.75E-5 & 5.99E-7 & -7.87E-7 & 3.64E-8 & -1.24E-10 & -2.55E-11 & -3.01E-11 & 2.89E-11 & -1.16E-12 & 0 & -5.99E-13 &  -9.08E-12\\
% 100*   &  -32.668 & 28.800 & 3.051 & 0.425 & 0.261 & -0.006 & 0.124 & 0.016 & 3.489E-4 & 2.678E-3 & 1.489E-3 &  -9.762E-6\\Models with (*) are from Paper I and include the entire VMS evolution through the MS and post-MS for comparison.

      \hline
    \end{tabular}
    \caption{Ejected masses (top) and wind yields (bottom) for our grid of models, calculated from the onset of core He-burning until core O-exhaustion (i.e. during the cWR phase). Initial masses, yields and ejected masses are provided in solar mass units. Comparison models are included for a pure Helium star with 40\Mdot\ (*) applying mass-loss rates from \citet{NugisLamers}, and for a post-VMS 32\Mdot\ ($\dagger$) model from \citet{Higgins+23} for which only the post-MS contribution is included here.}
    \label{tab:yields}
\end{table}
\end{landscape}

\subsection{cWR wind yields}
Stellar wind yields (Table \ref{tab:yields}, bottom) are a useful input for GCE models as they compare the enrichment of the host environment relative to the initial composition of the star. Therefore, positive chemical yields demonstrate enrichment of a given isotope while the negative yields show the removal of a given isotope relative to the initial composition. We find that all cWR models yield positive amounts of $^{14}$N, $^{23}$Na, $^{26}$Mg, $^{26}$Al, and $^{27}$Al. Simultaneously all models provide negative yields of $^{1}$H, and $^{20}$Ne. The most massive cWR stars (20 $<$ M/\Mdot\ $<$ 50) also yield positive amounts of $^{12}$C, $^{16}$O, $^{19}$F, and $^{22}$Ne ($>$25\Mdot). The key products of core H-burning, which are also released via winds during core He-burning are $^{14}$N,  $^{23}$Na, $^{26,27}$Al, and $^{28}$Si. The main He-burning products in our wind yields are $^{12}$C, $^{16}$O, $^{22}$Ne, and $^{26}$Mg.

\begin{figure}
    \centering
    \includegraphics[width = \columnwidth]{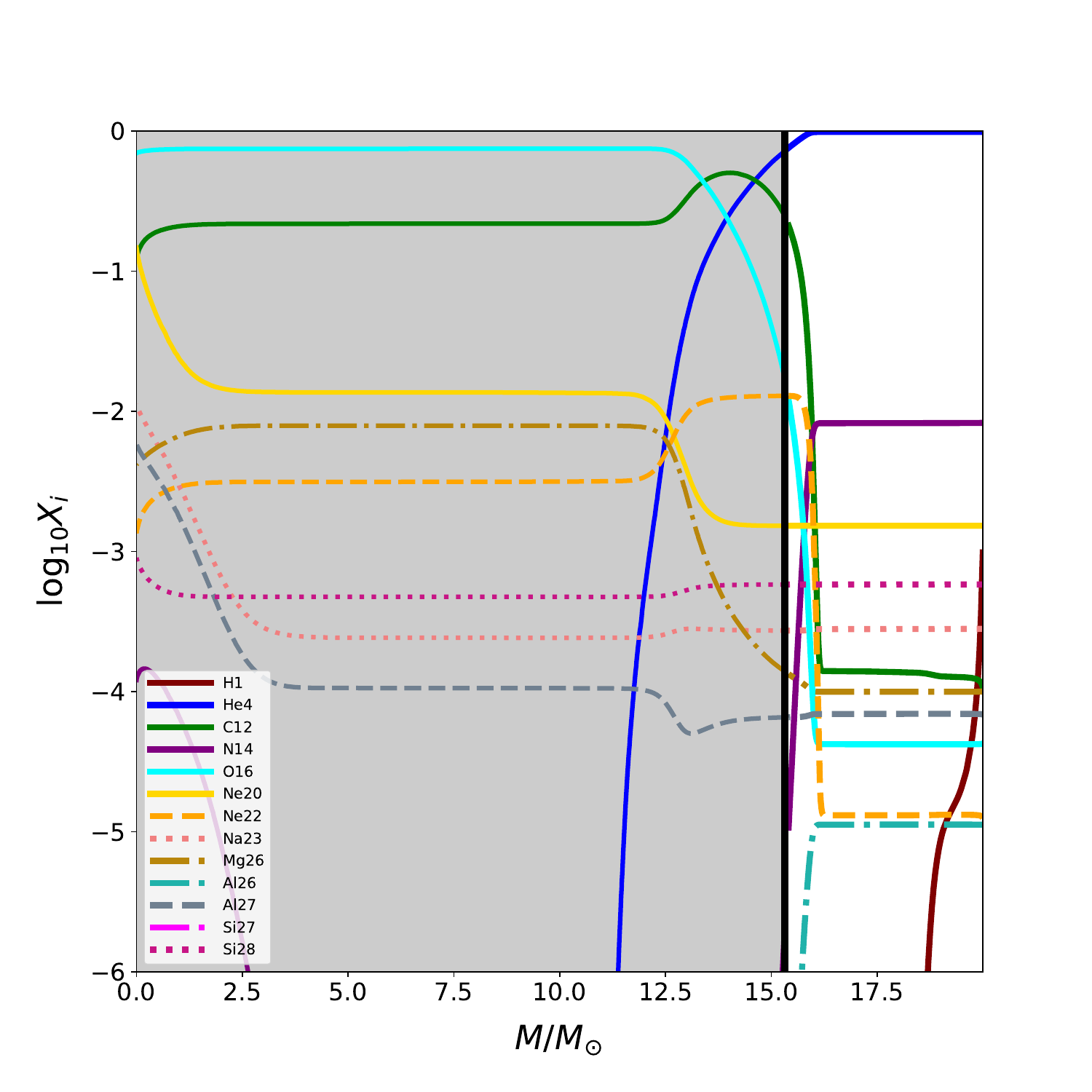}
    \caption{Time evolution of surface isotopes in mass fractions as a function of stellar mass during core He-burning of a 20\Mdot\ Helium star. As the star loses mass through stellar winds, the surface abundances evolve right to left. The grey shaded region illustrates the final mass after core He-burning.}
    \label{fig:20WR}
\end{figure}

\begin{figure}
    \centering
    \includegraphics[width = \columnwidth]{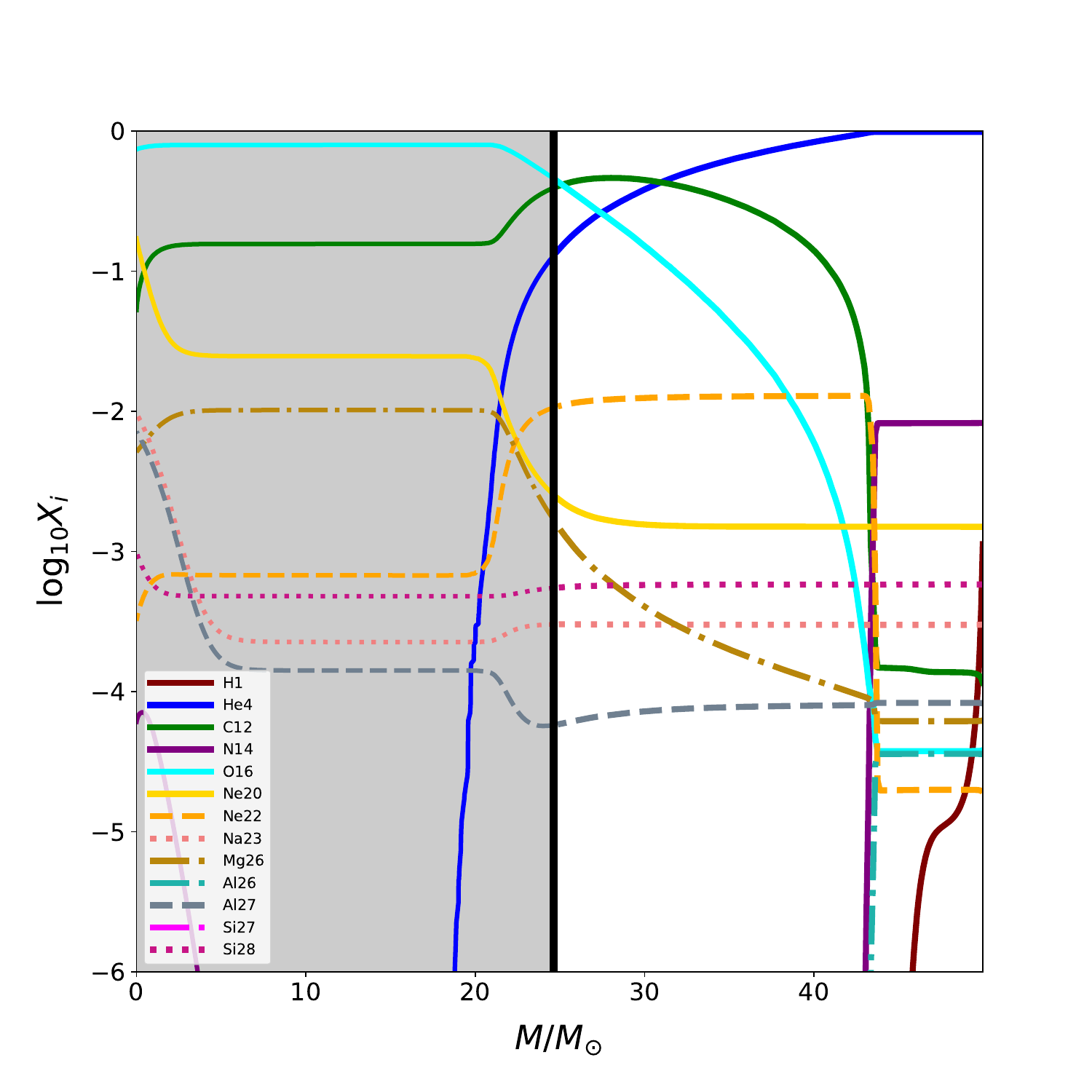}
    \caption{Time evolution of surface isotopes in mass fractions as a function of stellar mass during core He-burning of a 50\Mdot\ Helium star. The grey shaded region illustrates the mass left in the star after core He-burning.}
    \label{fig:50WR}
\end{figure}

We note that all models eject increasing amounts of each isotope with increasing stellar mass due to the luminosity-dependency of cWR winds. We illustrate the ejected mass of each isotope for a 20\Mdot\ star in Fig. \ref{fig:20WR} where the surface evolution of each isotope is shown from right to left in the white region, while the final He-exhausted core is shown in grey. Fig. \ref{fig:20WR} highlights the dominant ejecta which are $^{4}$He and $^{14}$N, with a smaller fraction of $^{20}$Ne, $^{23}$Na and $^{28}$Si. This 20\Mdot\ star remains N-rich at the surface throughout core He and C-burning, losing only $\sim$ 5\Mdot\ during the WR stage. Comparatively, the surface evolution of a 50\Mdot\ cWR is shown in Fig. \ref{fig:50WR} where a significant portion of the star's mass has been lost through stellar winds, with 50\% of the mass retained in the He-exhausted core (grey). We notice that the N-rich layer is stripped quickly, revealing the C-rich He-fusion products at the surface, and spending most of the stars cWR phase as a WC star. Towards the end of the stars evolution, the 50\Mdot\ cWR enriches in $^{16}$O at the surface. \cite{mm12} similarly find that in order for cWR stars to eject measurable amounts of He-burning products (i.e.$^{12}$C, $^{16}$O), the WC phase is crucial. Therefore, the yields of $^{12}$C and $^{16}$O are most significant at the highest mass ranges ($\sim$ 30-50\Mdot). We find that the yields for these isotopes increase notably by a factor of 2-4 at this mass range ($\geq$ 30\Mdot).

Interestingly, the Ne isotope which accompanies the C-rich phase in the 50\Mdot\ model, is the isotope $^{22}$Ne rather than the $^{20}$Ne which was most abundant in the 20\Mdot\ surface evolution. The $^{22}$Ne abundance dramatically increases as $^{14}$N is depleted due to 2 $\alpha$ captures which almost instantaneously converts the high $^{14}$N abundance to $^{22}$Ne, at the start of He-burning. More massive cWR stars will eject more $^{22}$Ne than $^{20}$Ne since they eject the $\alpha$-processed $^{22}$Ne during the C-rich phase rather than large quantities of $^{14}$N. This also has consequences for the remaining $^{22}$Ne and neutron source for the weak s-process, discussed in Sect. \ref{sprocess}.

The $^{22}$Ne/$^{20}$Ne ratio has been observed to be much higher in cosmic rays in the Milky Way than in the solar system \citep{garciamunoz79, weidenbeck81, lukasiak94, binns01}. The stellar winds of the most massive cWR stars are considered to eject significant quantities of Ne isotopes, while also forming superbubbles and supernovae, predicted to be the source of cosmic rays detected in the Milky Way  \citep{HigLing03}. Moreover, these superbubbles are proposed to be enriched not only by the resulting supernovae but by the vast amount of $^{22}$Ne ejected by cWR winds \citep{linghig00}. The important role that cWR stars may play in determining the solar Ne ratios has been further explored by \cite{binns05}. Therefore, the Ne yields of cWR winds may be key to better understand the Galactic $^{22}$Ne/$^{20}$Ne ratio.

\begin{figure}
    \centering
    \includegraphics[width = \columnwidth]{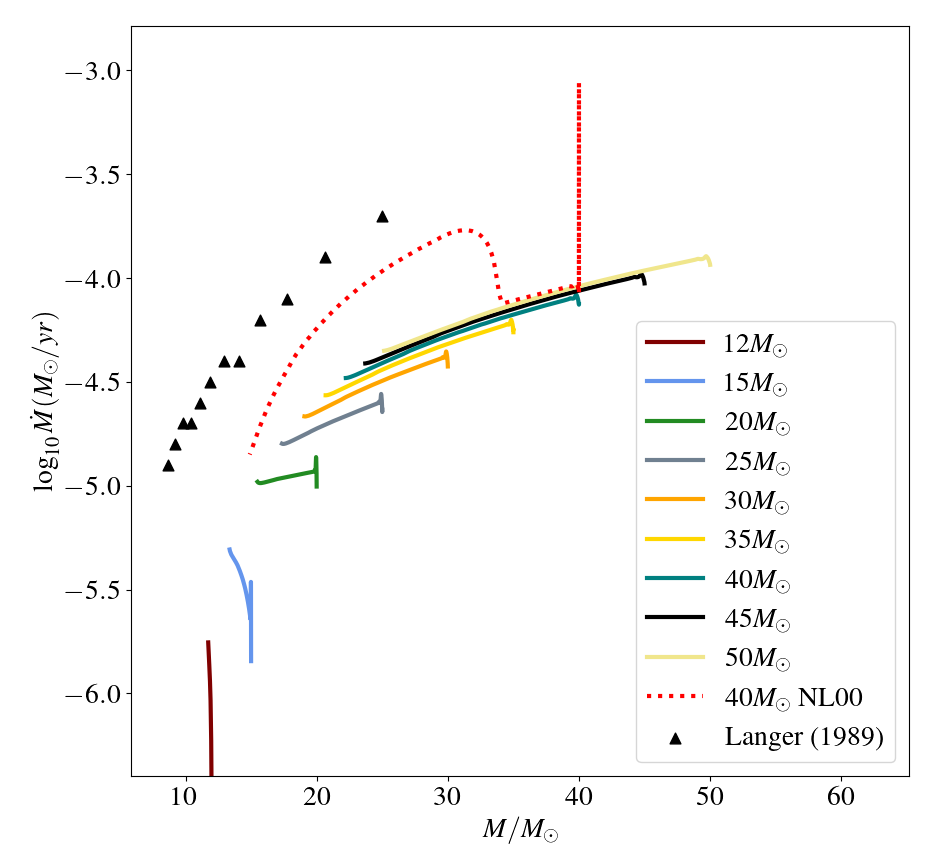}
    \caption{Mass-loss rates as a function of mass for our model grid (12-50\Mdot) shown in solid coloured lines. A 40\Mdot\ model applying rates from \citet{NugisLamers} is shown (dashed line) representing the 40$^{*}$\Mdot\ model from Table \ref{tab:yields}. The mass-dependent rates from \citet{langer89}, included by \citet{meynetarnould00} are illustrated by black triangles.}
    \label{fig:mdot}
\end{figure}

% \subsection{Comparison with literature}
Previously, stellar evolution models of cWR stars have implemented wind rates from \cite{NugisLamers}, applied to stars with surface H $<$ 0.4 based on empirical results from WR stars at \Zdot. We calculate a test case for a high mass cWR model where the effects of wind mass loss will be most prominent. Table \ref{tab:yields} includes a 40\Mdot\ model (*) which applies the \cite{NugisLamers} wind prescription, as a comparison to our 40\Mdot\ model which applies the updated hydrodynamically-consistent rates from \cite{SV2020}, see Fig. \ref{fig:mdot}. We find a notable difference in final masses at the end of core O-burning, with 21.8\Mdot\ for our 40\Mdot\ model and 14.7\Mdot\ for the comparison model applying \cite{NugisLamers} rates. The wind yields which are predominantly affected are the He and C ejecta with an additional 4.6\Mdot\ and 2.3\Mdot\ lost with \cite{NugisLamers} rates respectively. We note that $^{19}$F and $^{22}$Ne yields also increase with higher mass-loss rates from \cite{NugisLamers}. Interestingly, the amount of $^{26}$Al is not affected by the choice of wind prescription, since these outer enriched layers are stripped quickly in both cases, and $^{26}$Al is not produced during core He-burning. This confirms that the core H-burning VMS are key sources of $^{26}$Al, and regardless of wind rates cWR stars do not yield significant amounts of  $^{26}$Al. 

\subsection{Production of $^{19}$F}\label{nuclear_HHe}
The origin of fluorine ($^{19}$F) is not well constrained in the solar neighbourhood \citep{Ryde20}. $^{19}$F is destroyed during core H and He burning via the reactions $^{19}$F(p, $\alpha$)$^{16}$O and $^{19}$F($\alpha$, p)$^{22}$Ne, so determining which sources can build up an observable reservoir of $^{19}$F is key for better understanding the observed $^{19}$F abundances \citep{spitoni}. Massive stars and their resulting cWR stars have been suggested to produce $^{19}$F and eject moderate yields of $^{19}$F before it is destroyed in further reactions \citep{meynetarnould00}. This production source has been further explored by \cite{cunha03, renda04, cunha08}, but is questioned by \cite{palacios05} as the yields predicted by their cWR models are significantly lower than that of \cite{meynetarnould00}. \cite{cunha03} suggest that cWRs can eject higher quantities of $^{19}$F, particularly at higher $Z$ ($\sim$ \Zdot). The contribution from asymptotic giant branch (AGB) stars has also been considered by \cite{olive19}, while the final nucleosynthesis at core-collapse in massive binary stars has been suggested to produce significant amounts of $^{19}$F by \cite{farmer23}.

During core H-burning there are lots of protons available, therefore many proton-capture reactions take place, and $^{19}$F can be produced as a continuation of CNO$_{\rm{II-III}}$ via \\
$^{14}$N(p,$\gamma$)$^{15}$O($\beta$$^{+}$)$^{15}$N(p,$\gamma$)$^{16}$O(p,$\gamma$)$^{17}$F, $^{17}$F($\beta$$^{+}$)$^{17}$O(p,$\gamma$)$^{18}$F($\beta$$^{+}$)$^{18}$O(p, $\gamma$)$^{19}$F.\\
However, during the CNO cycle, $^{19}$F is destroyed by $^{19}$F(p, $\alpha$)$^{16}$O and never reaches a high mass fraction at the surface to provide meaningful, or even positive net wind yields \citep{caughlanfowler88}, see also Figs. \ref{fig:20_Xs} and \ref{fig:50_Xs}. We confirm this with our net wind yields of $^{19}$F for M$_{\rm{i}}$ $\geq$ 80\Mdot\ from Paper I which are all negative. As the H-burning core mass decreases dramatically with strong mass-loss rates on the main sequence, the He-burning core becomes too small to be uncovered by winds. Therefore, with mainly $^{19}$F-deficient yields provided during core H-burning, the net wind yields over the stellar lifetime are negative for this initial mass range. Note that this also applies to stars which retain their H envelope since the early core Helium products ($^{19}$F) will not be present at the surface in sufficient quantities before being reprocessed. During core He-burning, if there is sufficient H remaining, proton-captures can still take place. But if the star is a stripped Helium star, this will not occur, and $\alpha$-capture is very efficient. At the onset of core He-burning, $^{14}$N captures two $\alpha$-particles to produce $^{22}$Ne: 
$^{14}$N($\alpha$,$\gamma$)$^{18}$F($\beta$)
$^{18}$O($\alpha$, $\gamma$)$^{22}$Ne.
%, while $^{15}$N $\alpha$-captures to $^{19}$F. 
If there are protons remaining, or produced via (n, p) reactions, at the start of core He-burning then the proton-rich environment will permit %$^{14}$N($\alpha$,$\gamma$)$^{18}$F($\beta$)
$^{18}$O(p, $\alpha$)$^{15}$N($\alpha$, $\gamma$)$^{19}$F
($\alpha$,p)$^{22}$Ne. 
%$^{18}$O(p,$\gamma$)$^{19}$F
If not, then $^{19}$F can still be produced by $^{15}$N($\alpha$, $\gamma$)$^{19}$F from the $^{15}$N left over at the end of H-burning.
% But in stripped Helium stars, the $\alpha$-rich core will produce $^{19}$F via $^{15}$N ($\alpha$,$\gamma$)$^{19}$F. 

The synthesis of $^{19}$F relies on abundant quantities of neutrons, protons and $^{14}$N, where the neutrons become available via the $^{13}$C($\alpha$, n)$^{16}$O reaction. 
%RH The $^{14}$N(n, p)$^{14}$C reaction
Then (n, p) reactions, the $^{14}$N(n, p)$^{14}$C reaction in particular, 
can occur, creating a source of protons for $^{18}$O(p, $\alpha$)$^{15}$N, which is faster than the $^{18}$O(p, $\gamma$)$^{19}$F reaction, 
%leading to
which is followed by
$^{15}$N($\alpha$, $\gamma$)$^{19}$F. While in our models we do not consider $^{14}$C reactions, we have conducted a test and find that the addition of this reaction increases the abundance of $^{19}$F from log -5.2 by 0.3 dex in mass fraction or $\sim$ 5\%, in line with results from \cite{meynetarnould00}, however our net yields are not significantly affected.

\begin{figure}
    \centering
    \includegraphics[width = \columnwidth]{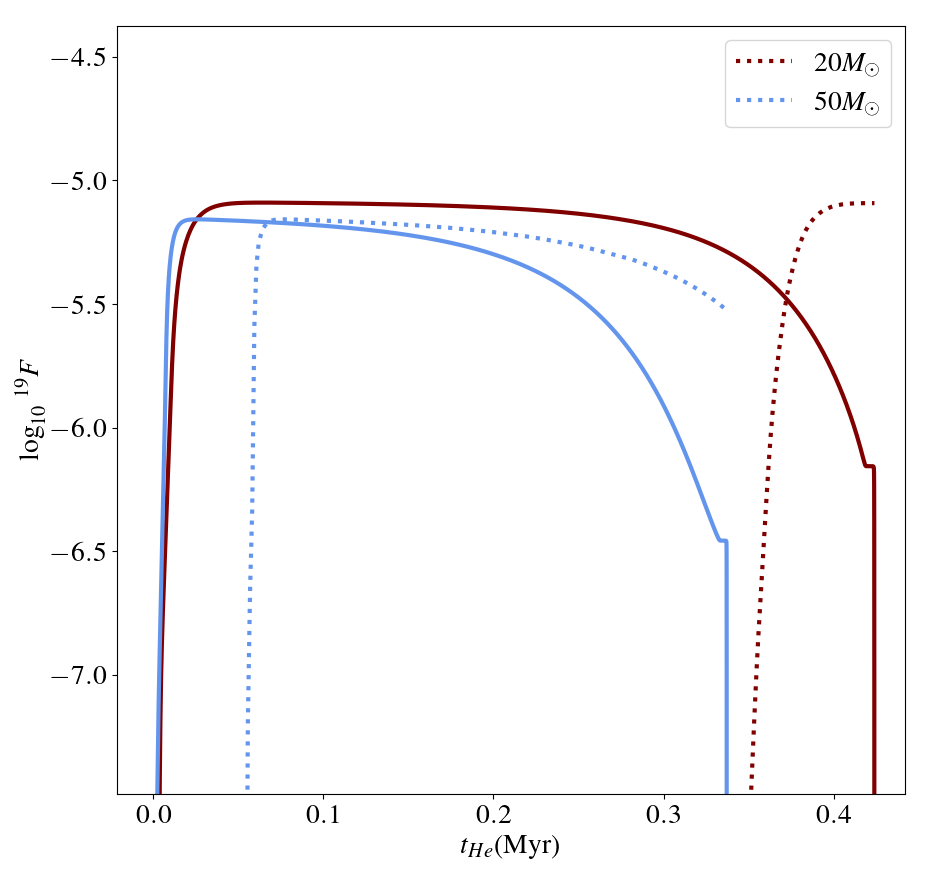}
    \caption{Surface evolution (dashed) and central (solid) abundance of $^{19}$F in 20\Mdot\ (red) and 50\Mdot\ (blue) models, as a function of core He-burning lifetime in Myrs.}
    \label{fig:f19}
\end{figure}
In the early stages of core He-burning, there is a build up of $^{19}$F, which dominates the $^{19}$F yields. Towards the end of core He-burning, $^{19}$F is destroyed by producing $^{22}$Ne. Therefore, if a star is stripped of its H envelope by the end of core H-burning, and can thereby start to expose He-burning products at the surface, then strong winds at the onset of core He-burning will lead to significant $^{19}$F wind yields.
% We find a correlation of faster $\alpha$-captures 
% which affect each of these reactions discussed 
% during core He-burning when compared to proton-capture reactions.
% , and we note similar trends in the central abundances of $^{18}$O, $^{14}$N and $^{15}$N. 
Interestingly, we find that our set of cWR models produce positive yields of $^{19}$F for masses greater than 20\Mdot\ ($\sim$10$^{-5}$\Mdot) relative to the initial composition
%RH
(the evolution of the surface composition for the 20 and 50\Mdot\ model is shown in Figs.\,\ref{fig:20_Xs} and 
 \ref{fig:50_Xs}, respectively). Figure \ref{fig:f19} illustrates that a 20\Mdot\ Helium star does not enrich in $^{19}$F at the surface until late in the core He-burning evolution ($\sim$ 0.35 Myrs), while a 50\Mdot\ star would already become enriched in $^{19}$F very early leading to significant $^{19}$F yields. The delay in $^{19}$F reaching the surface of a 20\Mdot\ star can be seen (red dashed line), compared to the negligible delay in $^{19}$F enrichment shown for a 50\Mdot\ star (blue dashed line).
%RH I would not write this: , throughout their entire evolution. 
This conclusion is in agreement with \cite{meynetarnould00}, which included even higher mass-loss rates from \cite{langer89} and the $^{14}$N(n, p)$^{14}$C reaction. While their models were evolved throughout the entire stellar evolution \citep[with high mass-loss rates from the H-ZAMS,][$\times$ 2]{dejager88}, thereby including the $^{19}$F-depleted material from the MS, by applying strong WR winds their models produce positive net $^{19}$F wind yields of $\sim$10$^{-4}$\Mdot. We note that \cite{palacios05} find reduced net yields ($\sim$10$^{-5}$ - $\sim$10$^{-6}$\Mdot) by adopting WR wind rates from \cite{NugisLamers} and updated NACRE reaction rates. However, our *40\Mdot\ test case with \cite{NugisLamers} wind rates from Table \ref{tab:yields} yields 4$\times$10$^{-5}$ more $^{19}$F than our comparable 40\Mdot\ model. Figure \ref{fig:mdot} demonstrates the higher mass-loss rates applied by \cite{meynetarnould00} and \cite{NugisLamers} in comparison to the updated rates by \cite{SV2020}. We conclude that while part of the core He-burning may occur in Nature before fully exposing the pure Helium core, our positive $^{19}$F yields of order 10$^{-5}$\Mdot\ highlight that pure Helium WR stars may in fact be an important source of $^{19}$F, through their winds.

\begin{figure}
    \centering
    \includegraphics[width = \columnwidth]{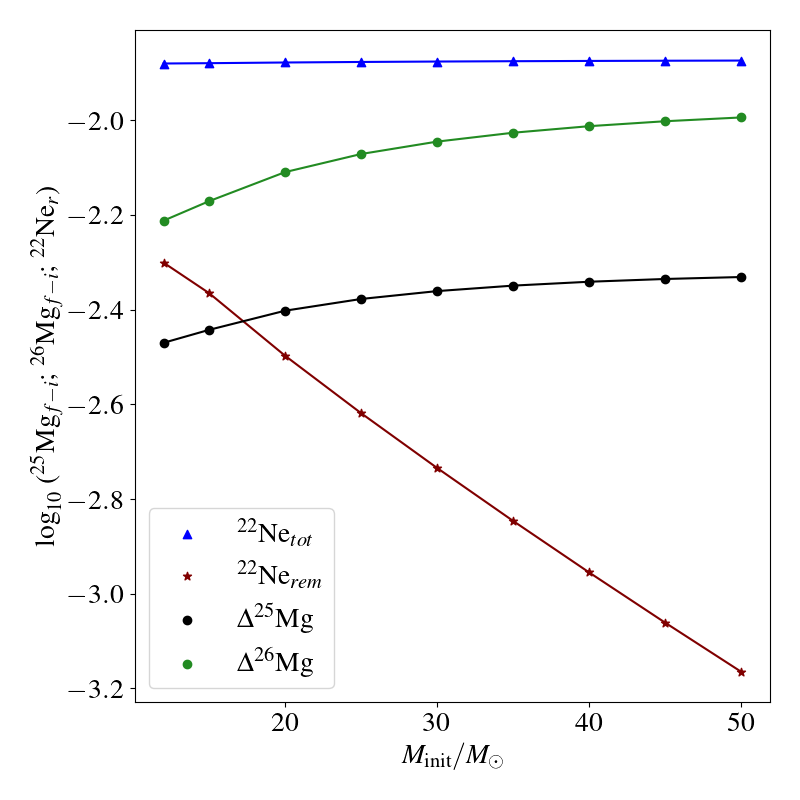}
    \caption{Amount of $^{25}$Mg or $^{26}$Mg synthesised (in mass fractions) during core He-burning (black dots, green dots), and remaining $^{22}$Ne (red stars) at He-exhaustion as a function of initial He-ZAMS mass of each model in our grid. The total $^{22}$Ne ($\Delta$$^{25}$Mg$_{\rm{f-i}}$ $\times$ 22/25 $+$ $\Delta$$^{26}$Mg$_{\rm{f-i}}$ $\times$ 22/26 $+$ $^{22}$Ne$_{\rm{r}}$) is shown with blue triangles.  }
    \label{fig:Ne22r}
\end{figure}

\section{Neutron source for weak s-process}\label{sprocess}
There is a rapid increase in $^{22}$Ne at the onset of He-burning due to the plentiful $^{14}$N from H-burning, (see the drop in $^{14}$N and rise in $^{22}$Ne at log $t - t_{f}$ $\sim$ 5.5 in Fig. \ref{fig:200_neut_Xc}). The $^{22}$Ne now $\alpha$-captures to $^{25}$Mg, ejecting a neutron each time. The $^{25}$Mg abundance increases by 3 orders of magnitude directly with the increase in $^{22}$Ne at He ignition, though then slowly increases during core He-burning (by another $\sim$2 orders of magnitude). This provides a substantial neutron source which enables the so called 
%RH updated end of this sentence
weak slow neutron-capture `s-process' where heavy elements beyond the iron (Fe) group are produced in hydrostatic stellar cores of massive stars \citep{Frisch16}. 
% The released neutrons in He-burning regions can then be captured by iron seeds and produce s-process elements from strontium up to barium \citep{mm12}.

The weak s-process mainly occurs during core He and C-burning phases since the later core O and Ne phases evolve at much higher central temperatures which prevent heavier s-process isotopes from surviving photodisintegration. During core C-burning heavy isotopes from the initially high $Z$ abundances ($\sim$\Zdot) can be neutron `poisons' which capture the neutrons and lower the neutron flux, impeding the s-process from being efficient \cite{mm12}. Therefore, the weak s-process is mainly effective during core He-burning. For this reason we focus on the neutron source for the weak s-process during core He-burning only. In lower $Z$ environments, the reduced quantity of $^{22}$Ne and iron seeds lead to inefficient weak s-process reactions also during core He-burning. While there are fewer weak s-process `poisons', they become more relevant and hence the quantity of weak s-process elements is expected to decrease with $Z$. 
%RH added this sentence:
Rotation-induced mixing may, however, significantly boost the weak weak s-process at low metallicities \citep{Frisch16}.
\begin{figure*}
    \centering
    \includegraphics[width = \textwidth]{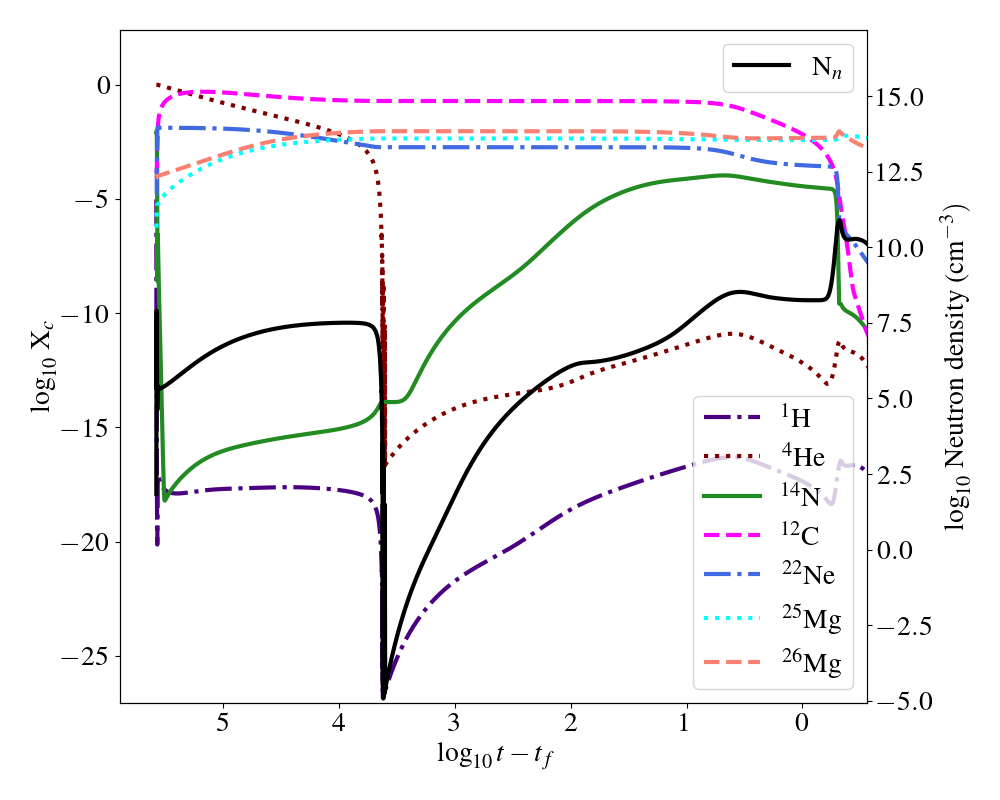}
    \caption{Evolution of the central composition (left axis) and central neutron density (right axis) in mass fractions with time in log-scale until core C-exhaustion for a 30\Mdot\ Helium star. }
    \label{fig:neut_Xc}
\end{figure*}

While the sequential $^{25}$Mg $\alpha$-capture to $^{28}$Si can occur, we find that this reaction is inefficient and has a negligible effect which does not lead to a notable destruction of $^{25}$Mg during the core He-burning phase. Therefore the relative difference in $^{25}$Mg (final - initial) can be an excellent proxy for the neutron exposure as this demonstrates how much of the $^{22}$Ne has been processed into $^{25}$Mg, releasing neutrons. The competing $^{22}$Ne($\alpha$, $\gamma$)$^{26}$Mg reaction also occurs during core He-burning, and reduces the efficiency of producing neutrons from $^{22}$Ne.
%RH removed reactions. 
At the onset of core He-burning, the ($\alpha$, $\gamma$)$^{26}$Mg reaction is more efficient (T$_{\rm{c}}$ $\sim$ 0.1-0.2GK, see Table \ref{tab:properties}), but for the remainder of core He-burning, the ($\alpha$, n)$^{25}$Mg reaction is dominant \citep{adsley21}.

Figure \ref{fig:Ne22r} demonstrates the efficiency of neutron production in the core as a function of stellar mass for our model grid via $^{22}$Ne($\alpha$, n)$^{25}$Mg, with $\Delta$ $^{25}$Mg (black dots) representing the final $^{25}$Mg abundance relative to the initial $^{25}$Mg, to illustrate the amount of $^{25}$Mg that has been synthesised during core He-burning. We also present the relative $\Delta$ $^{26}$Mg (green dots) which demonstrates how much $^{22}$Ne has been processed into $^{26}$Mg without producing neutrons. The amount of $^{22}$Ne remaining at the end of core He-burning (red stars) therefore represents the leftover $^{22}$Ne which has not been synthesised into $^{25}$Mg to produce neutrons yet, or into $^{26}$Mg. We find that the neutron production increases from 12-30\Mdot\ and plateaus at the highest mass range ($\sim$ 30-50\Mdot), while the remaining $^{22}$Ne shows a linear relation with increasing mass. The total $^{22}$Ne (synthesised to $^{25}$Mg or $^{26}$Mg, and $^{22}$Ne remaining) is presented for comparison (blue triangles). We confirm that the total $^{22}$Ne is constant with initial mass during core He-burning, relative to the total stellar mass (i.e. presented in mass fractions). For clarity, the $\Delta$ $^{25}$Mg (black), $\Delta$ $^{26}$Mg (green) and $^{22}$Ne$_{\rm{rem}}$ (red) equate to the total $^{22}$Ne (blue).

We find that models with higher initial masses (on the He-ZAMS) burn more $^{22}$Ne during core He-burning than lower mass models, leaving a lower abundance of $^{22}$Ne for the C-burning phase. The plateau seen in the abundance of $^{22}$Ne in Fig. \ref{fig:20WR} during core He-burning and at He-exhaustion provides the $\Delta$$^{22}$Ne, with the He-exhaustion abundance of $^{22}$Ne equating to the remaining $^{22}$Ne which has not been processed into $^{25}$Mg. Interestingly, for similar initial masses, the relative difference in $^{25}$Mg (representing the efficiency of the $^{22}$Ne-$^{25}$Mg reaction), and the amount of unprocessed $^{22}$Ne remaining, are on the same order of magnitude ($\sim$ 10$^{7}$ cm$^{-3}$) as stellar evolution theory \citep{Clay83} and are in agreement with the models from \cite{Frisch16}. 

Figure \ref{fig:20WR} shows a much lower 
%RH add surface
surface
abundance of $^{22}$Ne in a 20\Mdot\ star during the core He-burning stage (white region) in comparison to a 50\Mdot\ star (Fig. \ref{fig:50WR}). This illustrates that the subsequent plateau of $^{22}$Ne seen in the He-exhausted core (shaded region, $\sim$ 10\Mdot) of the 20\Mdot\ model in Fig. \ref{fig:20WR} is an order of magnitude higher than the plateau of $^{22}$Ne in the 50\Mdot\ model (Fig. \ref{fig:50WR}, $\sim$ 10\Mdot). The comparison between a 20\Mdot\ and 50\Mdot\ cWR star showcases that the main yields from the 20\Mdot\ model are H-processed isotopes, while the 50\Mdot\ model mainly ejects He-processed isotopes. Furthermore, the remaining central abundances (grey
region) of the 20\Mdot\ model illustrate a higher $^{22}$Ne abundance than in the corresponding 50\Mdot\ model because the central temperature
is lower in the 20\Mdot\ model and thus fewer $\alpha$-captures on $^{22}$Ne occur at the end of the core He-burning phase. 
% with reduced wind yields at lower masses, and reduced processing of $^{22}$Ne during the early He-burning phase, the neutron flux resulting from the $^{22}$Ne($\alpha$,n)$^{25}$Mg reaction would be higher with higher initial masses. 

We calculate the central neutron density by,
\begin{equation}
    N_{n} = \rho N_{A} n
\end{equation}
%RH fixed italics/straight fonts ; if you prefer straight fonts, you need to also use straight fonts in the formula above
where $n$ is the central neutron abundance in mass fraction, $N_{A}$ is Avogadro's number, and $\rho$ is the central density. Figure \ref{fig:neut_Xc} illustrates the central neutron density ($N_{{n}}$) and central composition with time until core C-exhaustion for a 30\Mdot\ cWR star. We note the sharp peak in $N_{{n}}$ at the beginning (log$_{10} t-t_{f}$ $\sim$ 5.5) due to the $^{13}$C($\alpha$, n) reaction. The prolonged increase in the N$_{\rm{n}}$ to 10$^{7.5}$ during core He-burning (5 $<$ log$_{10} t-t_{f}$ $<$ 4) shows the production of neutrons from $^{22}$Ne which is simultaneously decreasing, and the production of $^{25}$Mg which also increases at this point. We can see a second increase in the N$_{\rm{n}}$ during core C-burning (log$_{10} t-t_{f}$ $\sim$ 1) where $^{22}$Ne drops again. Since our simulations do not incorporate a complete s-process nuclear network, we do not trace the reprocessing of neutrons in the late phases of evolution (0 $<$ log$_{10} t-t_{f}$), but we will study the full weak s-process in a future work. We note that we have considered the neutron production, and not the neutron capture or destruction by Fe or other isotopes. A comparable central composition and neutron density plot is provided for a VMS with M$_{\rm{i}}$ $=$ 200\Mdot\ in Fig. \ref{fig:200_neut_Xc}, which illustrates both the core H and He-burning phases.

We find that the maximum %RH N$_{\rm{n}}$
$N_{{n}}$ during core He-burning is 3.21$\times$10$^{7}$cm$^{\rm{-3}}$ for a 30\Mdot\ stripped cWR model. Similarly, we find that a 32\Mdot\ post-VMS (M$_{\rm{H-ZAMS}}=$200\Mdot) cWR which is also stripped of H, has a maximum central N$_{\rm{n}}$ of  2.94$\times$10$^{7}$cm$^{\rm{-3}}$, which is comparable to models by \cite{Frisch16} (see their models A25s0 with $N_{{n}}$ $=$ 1.56$\times$10$^{7}$cm$^{\rm{-3}}$ and A40s4 with $N_{{n}}$ $=$ 1.42$\times$10$^{7}$cm$^{\rm{-3}}$). Since our models are pure stripped He stars which predict receding convective cores, they cannot grow by replenishing from a H-shell reservoir above the core. Comparably, the models by \cite{Frisch16} evolve as standard O supergiants with a H-shell above the He core, allowing a higher $\alpha$-source to generate the $^{22}$Ne-$^{25}$Mg reaction. It is interesting that while our pure Helium stars do not have an additional source of Helium to draw from, the maximum 
%RH N$_{\rm{n}}$
$N_{{n}}$ is very similar to the non-stripped He-burning models of \cite{Frisch16}. On the other hand, our stripped Helium models have the benefit of disregarding the stripping mechanism, and therefore provide chemical yields and conclusions which are applicable to both binary and single star channels alike. Finally, we find that the maximum central %RH N$_{\rm{n}}$
$N_{{n}}$ scales with initial mass (15-50\Mdot), as expected. However, we find that the growing core mass of our 12\Mdot\ star actually leads to the highest neutron density due to a higher central density and a dredge-down of Helium from the outermost layers.

% Therefore the increase in N$_{\rm{n}}$ during these final stages is not physical, and as mentioned above, the central temperature during core O-burning would be too high to permit stable s-process isotopes in this stage. 
\begin{figure}
    \centering
    \includegraphics[width = \columnwidth]{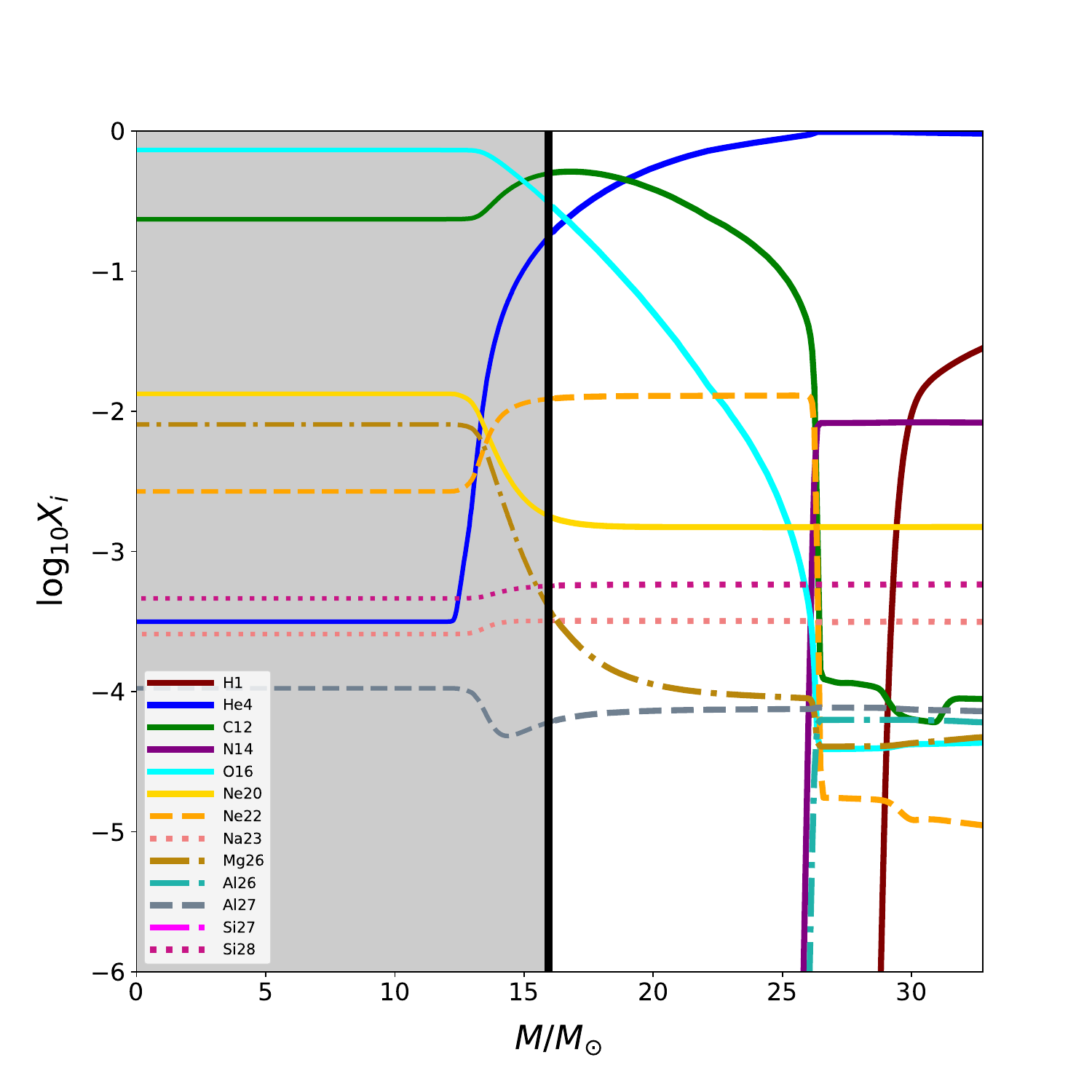}
    \caption{Time evolution of the surface composition during core He-burning in log-scale as a function of stellar mass with the interior composition shown at the end of core He-burning, for a model with an initial mass of 100\Mdot. The final interior composition at the end of core He-burning is shown in the grey shaded region (left) while the ejected material lost during the core He-burning phase can be seen in white (right).}
    \label{fig:yield100HeTams}
\end{figure}

\section{Comparison with VMS}\label{vms}
We explore the nucleosynthesis of cWR stars which have been evolved from the He-ZAMS, though follow the H-burning nucleosynthesis and omitting MS winds. The benefit of this method allows consideration of H-processed material which is then key for He-burning products. This includes the reservoir of $^{14}$N which is quickly processed into $^{22}$Ne, and later provides a source of neutrons for the weak s-process. While we do not consider how cWR stars are formed, our pure Helium models are relevant for a wide range of progenitor channels (via extreme rotation, VMS or binary stripping). We evolve a range of pure Helium stars from 12-50\Mdot\ to represent the variety of formation channels, where 50\Mdot\ is an upper limit for creating cWRs at \Zdot, comfortably encompassing observed WRs in the Galaxy, \citep{Crow07}.

In this section, we evaluate the contribution of cWR stars from the He-ZAMS, but utilise a stripped Helium star with its prior evolution history as a VMS from Paper I. In this case, a pure Helium star can begin burning He as an already exposed Helium core via strong VMS winds on the MS. We explore the consequences of this prior evolution, in comparison to our pure He-ZAMS models presented in this work. Finally, in this section we separate the main contributions from cWRs and VMS.

In Paper I, we provided ejected masses and wind yields of 50-500\Mdot\ stars from core H-burning until O-exhaustion. From \cite{Sabh22, Higgins22} we found that VMS (M$_{\rm{i}}$ $\geq$ 100\Mdot) lose substantial amounts of mass on the MS due to the optically-thick wind regime where stars above the transition point \citep{vink11, VG12} experience enhanced winds, leaving all TAMS masses converging to $\sim$ 32\Mdot, regardless of initial mass. \cite{Goswami} also present a range of stellar wind and supernovae yields, accounting for the IMF with M$_{\rm{i}}$ $<$ 350\Mdot, finding that VMS are crucial in reproducing the [O/Fe] ratios of thick-disk stars and the overall Galactic chemical enrichment.

We find that our cWR models eject similar amounts of $^{22}$Ne and $^{23}$Na when compared to VMS progenitors. Moreover, the 200\Mdot\ model ejects more $^{14,15}$N, $^{17,18}$O, $^{20,21}$Ne, $^{23}$Na, $^{24,25,26}$Mg, and $^{26,27}$Al than the 30\Mdot\ cWR star. On the other hand, the 30\Mdot\ Helium star ejects more $^{12}$C, $^{16}$O, and $^{22}$Ne than the 200\Mdot\ model.

% Therefore, massive cWR stars with M $>$ 30\Mdot\ likely evolve from stars below the transition point ($\sim$ 50-80\Mdot). 
In Paper I, we found that substantial amounts of $^{26}$Al were ejected by VMS on the MS as a result of enhanced stellar winds, while the post-MS resulted in $\sim$ 10$^{-2}$\Mdot\ of the decayed $^{26}$Mg and proton-captured $^{27}$Al. Our cWR models, eject an order of magnitude less $^{26}$Mg and $^{27}$Al when compared to VMS, and yield 2 orders of magnitude less ($\sim$ 10$^{-5}$\Mdot) $^{26}$Al. The significantly reduced yields of $^{26}$Al from cWR when compared to VMS suggest that cWR are not a key source of $^{26}$Al. 

 As a result of the core H-burning winds included in the 200\Mdot\ star from Paper I, the ejected H-products are much higher than that of the cWR (see their Table 4). Similarly, the increased $^{14}$N produced by VMS leads to an initially higher central $^{19}$F abundance than that of the stripped cWR stars. However, the net $^{19}$F yields for all VMS are negative (M$_{\rm{i}}$ $>$ 80\Mdot) since the majority of the material ejected is $^{19}$F-depleted. We compare the post-MS (He-burning until O-exhaustion) net yields of our 30\Mdot\ cWR model and a 32\Mdot\ post-VMS model in Table \ref{tab:yields}. Interestingly, the post-VMS model confirms that the evolutionary channel towards forming our pure Helium stars does not impact the net yields significantly. While the 32\Mdot\ model ejects slightly more $^{4}$He, $^{12}$C, $^{22}$Ne, $^{23}$Na and $^{26,27}$Al relative to its mass compared to our 30-35\Mdot\ cWRs, this is mainly due to the additional available protons during the MS evolution and the different wind prescription applied during core He-burning \citep{Sabh22}. We note that the $^{19}$F net yields are lower for the 32\Mdot\ model compared to the cWR models, since $\alpha$-captures are more efficient than proton-captures in the production of $^{19}$F during core He-burning. This confirms that the main source of $^{19}$F is not (very) massive stars, but exposed pure Helium stars which enrich quickly in $^{19}$F and eject it before it is destroyed. As long as VMS lose material in their winds which is enriched in H-burning products, they cannot enrich their surroundings with $^{19}$F. On the contrary, they eject $^{19}$F-depleted material. When the He-core is exposed sufficiently early during the core He-burning phase, their winds may then be enriched in $^{19}$F. Therefore, the net effect of their entire evolution will be positive or negative yields of $^{19}$F, depending on the importance of the mass-loss occurring during these two evolutionary stages.

We have compared the stellar parameters of the post-VMS evolved WR stars (from the onset of core He-burning) which all reached the He-ZAMS with M$=$ 32\Mdot, with the 30\Mdot\ cWR model presented in this work. We find that the $T_{\rm{eff}}$, luminosities, mass and surface abundances evolve very similarly, within 0.1dex. Furthermore, the central temperature evolution of both the cWR and post-VMS WR are highly comparable throughout the He-C-O burning phases. We note that the maximum neutron density discussed previously is also comparable in both models. We therefore find, that the evolutionary channel through which a stripped Helium star of a given mass forms has negligible effect on the stellar properties discussed in this work, and that the nucleosynthesis and stellar parameters are not significantly affected by the prior evolution. 

\section{Galactic WR observations}\label{obs}

Observations of cWR stars in the Milky Way, LMC and SMC have provided key insights into the progression between WR types (WN-WC-WO) and ultimately the resulting SNe types. \cite{Hamann06} analysed the observed Galactic WN sample with stellar atmosphere models providing stellar parameters, though with uncertain distances the luminosities were unconstrained. In \cite{Hamann+2019}, the updated GAIA distances provide improved accuracy in mass-loss rates and luminosities. Similarly, the observed Galactic WC sample were analysed by \cite{Sander19} to provide stellar parameters and wind properties of this evolved WR sequence, with a binary fraction of $\sim$ 40\% \citep{vanderHucht01}. Finally, the WC and WO stars were analysed by \cite{Tramper15} and later by \cite{aadland22} showing that with a few \% of surface O enrichment with a high surface C abundance, cWRs can be observed spectroscopically as a WO star. \cite{Crow07} provide further details on the observable surface properties of WR types (WN, WC). The observed WN abundances showcase elements which are processed by the CNO cycle (Fig. \ref{cno}) which lead to surface enrichments of $X_\text{N}$ $\sim$1\% by mass in observed Galactic WN stars, with negligible surface enrichment of $^{12}$C ($X_\text{C}$ $\sim$ 0.05\%). Galactic WC stars however, have been shown to present high enrichment of $^{12}$C with 10\% $<$ $X_\text{C}$ $<$ 60\%, and negligible surface $^{14}$N enrichment.

\begin{figure}
    \centering
    \includegraphics[width = \columnwidth]{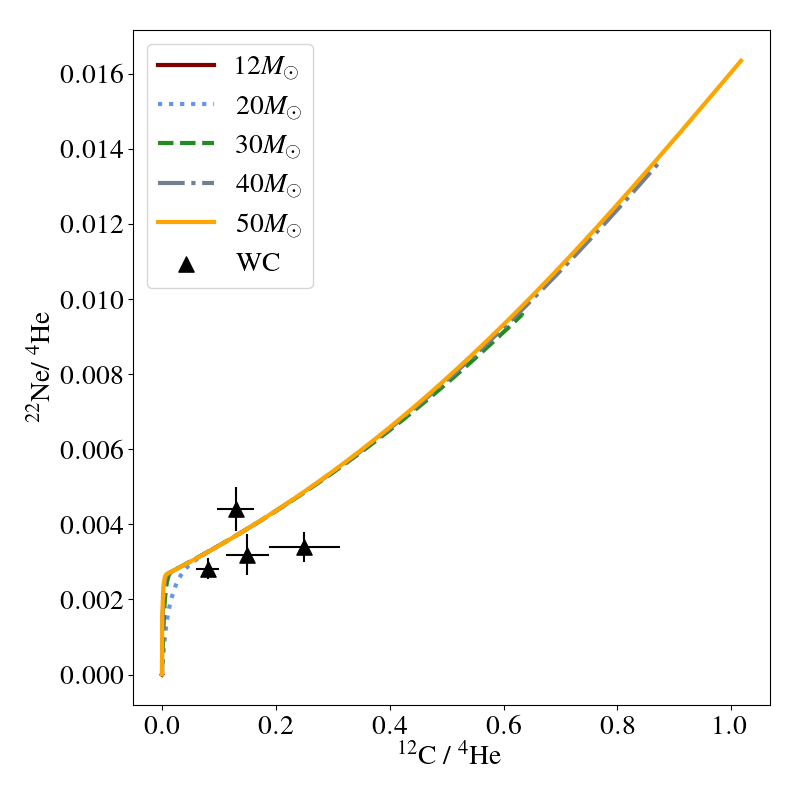}
    \caption{Surface ratios of Ne/He as a function of C/He by number for our grid of models (coloured lines) and observations of WC stars from \citet{Dessart2000} (black triangles).}
    \label{fig:CheNhe}
\end{figure}
\begin{figure}
    \centering
    \includegraphics[width = \columnwidth]{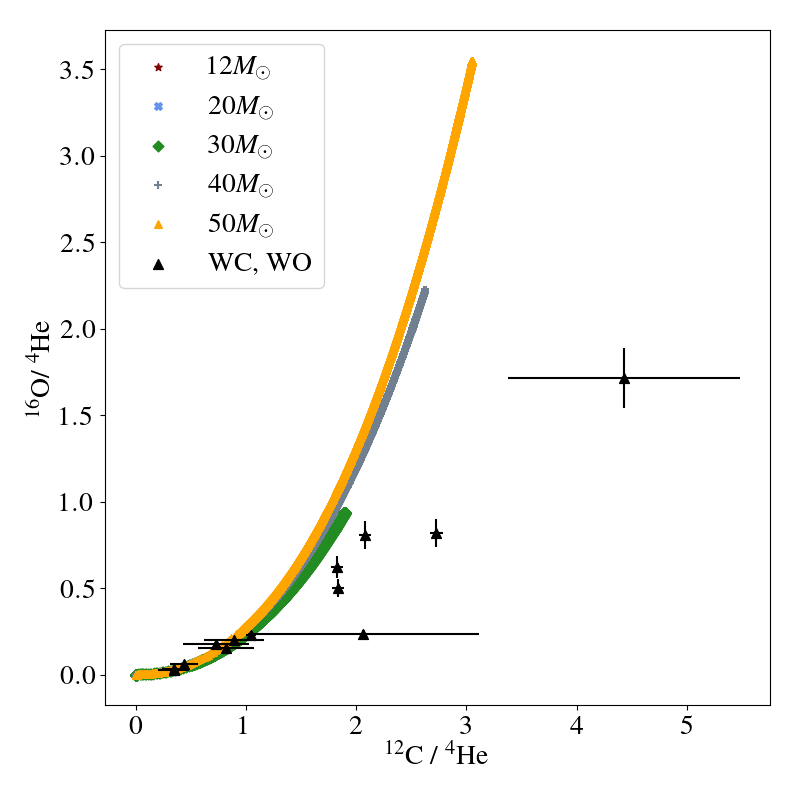}
    \caption{Surface abundance ratios of O/He as a function of C/He in mass fractions. Our grid of models are shown by the various coloured lines during core He-burning only. Observations of WC and WO stars from \citet{Tramper15} and \citet{aadland22} are shown by black triangles.}
    \label{fig:HeCO}
\end{figure}

We explore the 100\Mdot\ model from Paper I (comparable to the 200\Mdot\ model we discuss throughout this work) in Fig. \ref{fig:yield100HeTams} from the TAMS in more detail as a stripped He star. We identify the types of WR stars (WN, WC, WO) as a function of the core He-burning timescale and the evolving surface enrichment as mass loss peels off the outer layers exposing deeper fusion products. Initially, the N-rich WR star would be H-poor and He-rich with 10$^{-2}$ of $^{14}$N in mass fraction, presenting spectroscopically as a WN-type star (see $25<M/\Mdot\ <32$, Fig. \ref{fig:yield100HeTams}). At this point (M $\sim$ 25\Mdot) the $^{14}$N drops significantly at the expense of $^{22}$Ne, and the He-processed $^{12}$C is exposed at the stellar surface with an abundance of 10$^{-1}$ in mass fraction. This stage would correspond to the WC-stage of WR evolution and remains so with $^{12}$C as the dominant surface isotope (except for He) until the end of core He-burning. By peering into the He-exhausted core (grey shaded region), we can see that $^{16}$O quickly becomes the most abundant isotope, suggesting that a stripped WR star like that of Fig. \ref{fig:yield100HeTams} would only present spectroscopically as a WO star after core He-burning, with even shorter timescales ($\sim$ 1000 years). 
% Therefore, from VMS, we expect $\sim$ 1/3 of cWR stars to be WN-type and the remaining 2/3 to be WC-type. Post-He-burning stars might then provide WO-stars. 
From these results we can infer that WC stars must be late He-burning and post-He burning objects as the N-rich layer will not have been stripped during the early core He-burning stage, though this would also be a function of cWR winds. We provide further analysis of these results in Higgins et al. (in prep.).

We compare our stellar models with observed WC stars from \cite{Dessart2000} in Fig.\ref{fig:CheNhe} finding a good agreement between the observed [Ne/He] and [C/He] ratios, and our cWR model grid. Interestingly, since the $^{22}$Ne is produced from the CNO-processed $^{14}$N, this figure can act as a proxy of the initial CNO content \citep{meynet08}. The surface abundances of our cWR models do not change significantly during the first $\sim$ 70-80\% of the core He-burning timescale in the lower mass range (12-30\Mdot) of WR evolution, see also Fig. \ref{fig:20_Xs}. Similarly, the remaining $\sim$ 20\% of the core He-burning timescale in higher mass (30-50\Mdot) WR evolution does not show meaningful changes in the surface abundance, see Fig. \ref{fig:50_Xs}. 
% However, since $\sim$50\Mdot\ models show quick enrichment with $^{12}$C, we conclude that while all models are in good agreement with the observed WC ratios, lower mass models best reproduce the surface abundances of Galactic WC stars. 
The evolutionary trend and agreement with observations also align very clearly with that of \cite{Dessart2000}, see their Fig. 7.

We map the surface evolution of $^{12}$C and $^{16}$O as a function of $^{4}$He in Fig. \ref{fig:HeCO} with the observed abundances of WC and WO stars from \cite{Tramper15} and \cite{aadland22}. Our models are in good agreement with the late WC and WO stars for moderate [C/He] ratios ($\leq$2) which lie along the evolutionary tracks during the core He-burning phase. We present the core He-burning phase only for our model grid, but note that as previously discussed the surface abundances do not change significantly in the early (low mass) or late (high mass) phases of evolution. 
% This conclusion agrees well with the model predictions of \cite{MMHirschi}, where they similarly find weaker composition gradients in WC stars which do not showcase the core's evolution, but do agree with observations. 
Therefore, WC stars show abundances which are representative of partial He-burning, rather than the current central burning phase and as such leaves uncertainty about exactly which evolutionary stage WC stars are in. However, we conclude that the highest mass models (30-50\Mdot) reach higher [C/He] and [O/He] ratios towards the end of core He-burning. 
% Therefore, if the observed WC and WO stars are towards the end of core He-burning they may be best represented by the lower mass range (12-30\Mdot). 
It appears that from surface abundances alone, we infer that the observed WC and WO stars remain moderately enriched in $^{12}$C and $^{16}$O as a function of $^{4}$He and may not be evolved beyond core He-burning. The evolution of [C/He] and [O/He] ratios from our cWR models and with observed data align with that of \cite{aadland22}, see their Fig. 12.

\section{Conclusions}\label{conclusions}
In this work, we provide stellar wind yields for cWR stripped Helium stars with initial masses of 12-50\Mdot, implementing a large nuclear reaction network and hydrodynamically-consistent cWR winds from \cite{SV2020}. We compare the nucleosynthesis and wind yields of cWRs to that of VMS. The nucleosynthesis of isotopes such as $^{12}$C, $^{14}$N, $^{16, 18}$O, and $^{19}$F are traced as well as the $^{22}$Ne($\alpha$, n)$^{25}$Mg reaction which is the crucial neutron source for the weak s-process in massive stars at \Zdot. We calculate the maximum central neutron density (N$_{\rm{n}}$) for a range of masses, and compare with literature. Finally, we present a comparison of our $^{12}$C, $^{16}$O and $^{22}$Ne surface abundances with observed Galactic WR stars. We outline our main conclusions below.
\begin{itemize}
\item {We find that 12-20\Mdot\ cWR stars eject negligible amounts of each isotope in their winds, while 40-50\Mdot\ models eject significantly higher masses of $^{16}$O and $^{22}$Ne, as well as $^{26}$Mg and $^{27}$Al ($\sim$ 10$^{-3}$ \Mdot). }\\
\item {When compared to the ejected masses from VMS (with post-MS masses of 32\Mdot) in Paper I, we find that our cWR models (see 30\Mdot\ yields for direct comparison)} eject more $^{12}$C and $^{16}$O than our VMS models during their entire evolution, similar masses of $^{22}$Ne, $^{26}$Mg and $^{28}$Si, and less $^{26}$Al, $^{20}$Ne, $^{23}$Na.\\
\item {A 20\Mdot\ cWR star does not strip its outer layers sufficiently to become enriched with $^{12}$C at their surface, and as a result does not reach the WC stage during core He-burning. Since the later evolutionary stages are so short, the mass lost in these phases would not be enough to further strip the star to expose the C or O to produce WC/WO stars. Therefore, from 20\Mdot\ cWR stars, mostly WN stars would be produced. On the other hand, we find that a 50\Mdot\ star loses half of its mass during core He-burning and quickly enriches with $^{12}$C, thereby producing WC-type stars. }\\
\item {The observed [Ne/He] and [C/He] ratios of WC stars from \cite{Dessart2000} are well reproduced by our cWR model grid. Similarly, our cWR models produce [C/He] and [O/He] ratios which are in agreement with the observed WC and WO stars (for moderate [C/He] ratios $\leq$2) from \cite{Tramper15} and \cite{aadland22}. }\\
\item{We find comparable maximum central neutron densities during core He-burning for both the 30\Mdot\ cWR and 32\Mdot\ post-VMS }Helium stars, and show that they are in agreement with previous simulations of stars within comparable mass ranges.\\

\item{We find that Helium star models with M $>$ 20\Mdot\ yield positive amounts of $^{19}$F ($\sim$10$^{-5}$\Mdot) since their exposed cores can eject large quantities of $^{19}$F early in core He-burning before being reprocessed, illustrating the importance of Helium stars in enriching their host environments with $^{19}$F when their H envelope is removed by the onset of core He-burning.} \\
\item{ Interestingly, the formation channel towards forming pure Helium stars does not impact the subsequent internal structure or surface properties (luminosity or effective temperature). We find that by comparing post-VMS Helium stars from Paper I and cWR stars from this study, there are negligible differences in the composition and stellar properties from both evolutionary channels. We note that the remaining protons ($^{1}$H), and $^{14}$N present at the onset of core He-burning in post-VMS, have an effect on the reaction flow leading to $^{19}$F, via the $^{18}$O (p, $\alpha$) $^{15}$N($\alpha$, $\gamma$) $^{19}$F reactions. We note this difference in reaction flows between a post-VMS Helium star with 32\Mdot\ and a 30\Mdot\ cWR, but confirm that the overall total production of $^{19}$F is very similar.}\\
\item{Similarly, we find that the Helium star models presented in this work are independent of their formation channel either through binary stripping or single star evolution, and therefore can be implemented in GCE or population synthesis models without the assumption of how the Helium star lost its envelope.}\\
% Add conclusions of specific yields : F19, O16, C12, 
% Add comparison with VMS yields, comparison of key features (negligible differences in stellar parameters)
% Add conclusions on s-process, n densities
% Add final paragraph on usability of He star grid yields in wider GCE, pop synth without the preconception of how the He star lost its envelope either through single or binary stripping.
\end{itemize}

\section*{Acknowledgements}
The authors would like to thank the referee, Georges Meynet, for his expertise and detailed guidance which was significant in enhancing the content of the manuscript. We acknowledge MESA authors and developers for their continued revisions and public accessibility of the code. JSV and ERH are supported by STFC funding under grant number ST/V000233/1 in the context of the BRIDGCE UK Network. RH acknowledges support from STFC, the World Premier International Research Centre Initiative (WPI Initiative), MEXT, Japan and the IReNA AccelNet Network of Networks (National Science Foundation, Grant No. OISE-1927130). 
AML acknowledges support from STFC funding under grant number ST/V000233 in the context of the BRIDGCE UK Network. This article is based upon work from the ChETEC COST Action (CA16117) and the European Union’s Horizon 2020 research and innovation programme (ChETEC-INFRA, Grant No. 101008324). AACS is supported by the Deutsche Forschungsgemeinschaft (DFG - German Research Foundation) in the form of an Emmy Noether Research Group -- Project-ID 445674056 (SA4064/1-1, PI Sander) and acknowledges funding from the Federal Ministry of Education and Research (BMBF) and the Baden-Württemberg Ministry of Science as part of the Excellence Strategy of the German Federal and State Governments.
%%%%%%%%%%%%%%%%%%%%%%%%%%%%%%%%%%%%%%%%%%%%%%%%%%
\section*{Data Availability}
The data underlying this article will be shared on reasonable request
to the corresponding author.
\typeout{}
\bibliographystyle{mnras}
\bibliography{newdiff2.bib}

\begin{thebibliography}{}
\makeatletter
\relax
\def\mn@urlcharsother{\let\do\@makeother \do\$\do\&\do\#\do\^\do\_\do\%\do\~}
\def\mn@doi{\begingroup\mn@urlcharsother \@ifnextchar [ {\mn@doi@} {\mn@doi@[]}}
\def\mn@doi@[#1]#2{\def\@tempa{#1}\ifx\@tempa\@empty \href {http://dx.doi.org/#2} {doi:#2}\else \href {http://dx.doi.org/#2} {#1}\fi \endgroup}
\def\mn@eprint#1#2{\mn@eprint@#1:#2::\@nil}
\def\mn@eprint@arXiv#1{\href {http://arxiv.org/abs/#1} {{\tt arXiv:#1}}}
\def\mn@eprint@dblp#1{\href {http://dblp.uni-trier.de/rec/bibtex/#1.xml} {dblp:#1}}
\def\mn@eprint@#1:#2:#3:#4\@nil{\def\@tempa {#1}\def\@tempb {#2}\def\@tempc {#3}\ifx \@tempc \@empty \let \@tempc \@tempb \let \@tempb \@tempa \fi \ifx \@tempb \@empty \def\@tempb {arXiv}\fi \@ifundefined {mn@eprint@\@tempb}{\@tempb:\@tempc}{\expandafter \expandafter \csname mn@eprint@\@tempb\endcsname \expandafter{\@tempc}}}

\bibitem[\protect\citeauthoryear{{Aadland}, {Massey}, {Hillier}, {Morrell}, {Neugent}  \& {Eldridge}}{{Aadland} et~al.}{2022}]{aadland22}
{Aadland} E.,  {Massey} P.,  {Hillier} D.~J.,  {Morrell} N.~I.,  {Neugent} K.~F.,   {Eldridge} J.~J.,  2022, \mn@doi [\apj] {10.3847/1538-4357/ac66e7}, \href {https://ui.adsabs.harvard.edu/abs/2022ApJ...931..157A} {931, 157}

\bibitem[\protect\citeauthoryear{{Adsley} et~al.,}{{Adsley} et~al.}{2021}]{adsley21}
{Adsley} P.,  et~al., 2021, \mn@doi [\prc] {10.1103/PhysRevC.103.015805}, \href {https://ui.adsabs.harvard.edu/abs/2021PhRvC.103a5805A} {103, 015805}

\bibitem[\protect\citeauthoryear{{Arnett} \& {Thielemann}}{{Arnett} \& {Thielemann}}{1985}]{arnett85}
{Arnett} W.~D.,  {Thielemann} F.~K.,  1985, \mn@doi [\apj] {10.1086/163402}, \href {https://ui.adsabs.harvard.edu/abs/1985ApJ...295..589A} {295, 589}

\bibitem[\protect\citeauthoryear{{Arnett} et~al.,}{{Arnett} et~al.}{2019}]{arnett19}
{Arnett} W.~D.,  et~al., 2019, \mn@doi [\apj] {10.3847/1538-4357/ab21d9}, \href {https://ui.adsabs.harvard.edu/abs/2019ApJ...882...18A} {882, 18}

\bibitem[\protect\citeauthoryear{{Arnould}, {Paulus}  \& {Meynet}}{{Arnould} et~al.}{1997}]{Arnould97}
{Arnould} M.,  {Paulus} G.,   {Meynet} G.,  1997, \aap, \href {https://ui.adsabs.harvard.edu/abs/1997A&A...321..452A} {321, 452}

\bibitem[\protect\citeauthoryear{{Arnould}, {Goriely}  \& {Meynet}}{{Arnould} et~al.}{2006}]{arnould06}
{Arnould} M.,  {Goriely} S.,   {Meynet} G.,  2006, \mn@doi [\aap] {10.1051/0004-6361:20053966}, \href {https://ui.adsabs.harvard.edu/abs/2006A&A...453..653A} {453, 653}

\bibitem[\protect\citeauthoryear{{Asplund}, {Grevesse}, {Sauval}  \& {Scott}}{{Asplund} et~al.}{2009}]{asplund09}
{Asplund} M.,  {Grevesse} N.,  {Sauval} A.~J.,   {Scott} P.,  2009, \mn@doi [\araa] {10.1146/annurev.astro.46.060407.145222}, \href {https://ui.adsabs.harvard.edu/abs/2009ARA&A..47..481A} {47, 481}

\bibitem[\protect\citeauthoryear{{Binns} et~al.,}{{Binns} et~al.}{2001}]{binns01}
{Binns} W.~R.,  et~al., 2001, \mn@doi [Advances in Space Research] {10.1016/S0273-1177(01)00119-3}, \href {https://ui.adsabs.harvard.edu/abs/2001AdSpR..27..767B} {27, 767}

\bibitem[\protect\citeauthoryear{{Binns} et~al.,}{{Binns} et~al.}{2005}]{binns05}
{Binns} W.~R.,  et~al., 2005, \mn@doi [\apj] {10.1086/496959}, \href {https://ui.adsabs.harvard.edu/abs/2005ApJ...634..351B} {634, 351}

\bibitem[\protect\citeauthoryear{{Brinkman}, {Doherty}, {Pols}, {Li}, {C{\^o}t{\'e}}  \& {Lugaro}}{{Brinkman} et~al.}{2019}]{brinkman19}
{Brinkman} H.~E.,  {Doherty} C.~L.,  {Pols} O.~R.,  {Li} E.~T.,  {C{\^o}t{\'e}} B.,   {Lugaro} M.,  2019, \mn@doi [\apj] {10.3847/1538-4357/ab40ae}, \href {https://ui.adsabs.harvard.edu/abs/2019ApJ...884...38B} {884, 38}

\bibitem[\protect\citeauthoryear{{Caughlan} \& {Fowler}}{{Caughlan} \& {Fowler}}{1988}]{caughlanfowler88}
{Caughlan} G.~R.,  {Fowler} W.~A.,  1988, \mn@doi [Atomic Data and Nuclear Data Tables] {10.1016/0092-640X(88)90009-5}, \href {https://ui.adsabs.harvard.edu/abs/1988ADNDT..40..283C} {40, 283}

\bibitem[\protect\citeauthoryear{{Chieffi}, {Limongi}  \& {Straniero}}{{Chieffi} et~al.}{1998}]{chieffi98}
{Chieffi} A.,  {Limongi} M.,   {Straniero} O.,  1998, \mn@doi [\apj] {10.1086/305921}, \href {https://ui.adsabs.harvard.edu/abs/1998ApJ...502..737C} {502, 737}

\bibitem[\protect\citeauthoryear{{Clayton}}{{Clayton}}{1983}]{Clay83}
{Clayton} D.~D.,  1983, {Principles of stellar evolution and nucleosynthesis}

\bibitem[\protect\citeauthoryear{{Conti}, {Ebbets}, {Massey}  \& {Niemela}}{{Conti} et~al.}{1980}]{Conti80}
{Conti} P.~S.,  {Ebbets} D.,  {Massey} P.,   {Niemela} V.~S.,  1980, \mn@doi [\apj] {10.1086/157971}, \href {http://adsabs.harvard.edu/abs/1980ApJ...238..184C} {238, 184}

\bibitem[\protect\citeauthoryear{{Crowther}}{{Crowther}}{2007}]{Crow07}
{Crowther} P.~A.,  2007, \mn@doi [\araa] {10.1146/annurev.astro.45.051806.110615}, \href {https://ui.adsabs.harvard.edu/abs/2007ARA&A..45..177C} {45, 177}

\bibitem[\protect\citeauthoryear{{Crowther} \& {Walborn}}{{Crowther} \& {Walborn}}{2011}]{CrowWal11}
{Crowther} P.~A.,  {Walborn} N.~R.,  2011, \mn@doi [\mnras] {10.1111/j.1365-2966.2011.19129.x}, \href {https://ui.adsabs.harvard.edu/abs/2011MNRAS.416.1311C} {416, 1311}

\bibitem[\protect\citeauthoryear{{Cunha}, {Smith}, {Lambert}  \& {Hinkle}}{{Cunha} et~al.}{2003}]{cunha03}
{Cunha} K.,  {Smith} V.~V.,  {Lambert} D.~L.,   {Hinkle} K.~H.,  2003, \mn@doi [\aj] {10.1086/377023}, \href {https://ui.adsabs.harvard.edu/abs/2003AJ....126.1305C} {126, 1305}

\bibitem[\protect\citeauthoryear{{Cunha}, {Smith}  \& {Gibson}}{{Cunha} et~al.}{2008}]{cunha08}
{Cunha} K.,  {Smith} V.~V.,   {Gibson} B.~K.,  2008, \mn@doi [\apjl] {10.1086/588816}, \href {https://ui.adsabs.harvard.edu/abs/2008ApJ...679L..17C} {679, L17}

\bibitem[\protect\citeauthoryear{Cyburt et~al.,}{Cyburt et~al.}{2010}]{Cyburt10}
Cyburt R.~H.,  et~al., 2010, The Astrophysical Journal Supplement Series, 189, 240

\bibitem[\protect\citeauthoryear{{Dessart}, {Crowther}, {Hillier}, {Willis}, {Morris}  \& {van der Hucht}}{{Dessart} et~al.}{2000}]{Dessart2000}
{Dessart} L.,  {Crowther} P.~A.,  {Hillier} D.~J.,  {Willis} A.~J.,  {Morris} P.~W.,   {van der Hucht} K.~A.,  2000, \mn@doi [\mnras] {10.1046/j.1365-8711.2000.03399.x}, \href {https://ui.adsabs.harvard.edu/abs/2000MNRAS.315..407D} {315, 407}

\bibitem[\protect\citeauthoryear{{Eldridge} \& {Vink}}{{Eldridge} \& {Vink}}{2006}]{EV2006}
{Eldridge} J.~J.,  {Vink} J.~S.,  2006, \mn@doi [\aap] {10.1051/0004-6361:20065001}, \href {https://ui.adsabs.harvard.edu/abs/2006A&A...452..295E} {452, 295}

\bibitem[\protect\citeauthoryear{{Farmer}, {Renzo}, {de Mink}, {Marchant}  \& {Justham}}{{Farmer} et~al.}{2019}]{Farmer+2019}
{Farmer} R.,  {Renzo} M.,  {de Mink} S.~E.,  {Marchant} P.,   {Justham} S.,  2019, \mn@doi [\apj] {10.3847/1538-4357/ab518b}, \href {https://ui.adsabs.harvard.edu/abs/2019ApJ...887...53F} {887, 53}

\bibitem[\protect\citeauthoryear{{Farmer}, {Laplace}, {Ma}, {de Mink}  \& {Justham}}{{Farmer} et~al.}{2023}]{farmer23}
{Farmer} R.,  {Laplace} E.,  {Ma} J.-z.,  {de Mink} S.~E.,   {Justham} S.,  2023, \mn@doi [\apj] {10.3847/1538-4357/acc315}, \href {https://ui.adsabs.harvard.edu/abs/2023ApJ...948..111F} {948, 111}

\bibitem[\protect\citeauthoryear{{Freytag}, {Ludwig}  \& {Steffen}}{{Freytag} et~al.}{1996}]{Freytag1996}
{Freytag} B.,  {Ludwig} H.~G.,   {Steffen} M.,  1996, \aap, \href {https://ui.adsabs.harvard.edu/#abs/1996A&A...313..497F} {313, 497}

\bibitem[\protect\citeauthoryear{{Frischknecht} et~al.,}{{Frischknecht} et~al.}{2016}]{Frisch16}
{Frischknecht} U.,  et~al., 2016, \mn@doi [\mnras] {10.1093/mnras/stv2723}, \href {https://ui.adsabs.harvard.edu/abs/2016MNRAS.456.1803F} {456, 1803}

\bibitem[\protect\citeauthoryear{{Fujimoto}, {Krumholz}  \& {Tachibana}}{{Fujimoto} et~al.}{2018}]{Fuji18}
{Fujimoto} Y.,  {Krumholz} M.~R.,   {Tachibana} S.,  2018, \mn@doi [\mnras] {10.1093/mnras/sty2132}, \href {https://ui.adsabs.harvard.edu/abs/2018MNRAS.480.4025F} {480, 4025}

\bibitem[\protect\citeauthoryear{{Gaidos}, {Krot}, {Williams}  \& {Raymond}}{{Gaidos} et~al.}{2009}]{gaidos09}
{Gaidos} E.,  {Krot} A.~N.,  {Williams} J.~P.,   {Raymond} S.~N.,  2009, \mn@doi [\apj] {10.1088/0004-637X/696/2/1854}, \href {https://ui.adsabs.harvard.edu/abs/2009ApJ...696.1854G} {696, 1854}

\bibitem[\protect\citeauthoryear{{Garcia-Munoz}, {Simpson}  \& {Wefel}}{{Garcia-Munoz} et~al.}{1979}]{garciamunoz79}
{Garcia-Munoz} M.,  {Simpson} J.~A.,   {Wefel} J.~P.,  1979, \mn@doi [\apjl] {10.1086/183043}, \href {https://ui.adsabs.harvard.edu/abs/1979ApJ...232L..95G} {232, L95}

\bibitem[\protect\citeauthoryear{{Gilkis}, {Vink}, {Eldridge}  \& {Tout}}{{Gilkis} et~al.}{2019}]{Gilkis+2019}
{Gilkis} A.,  {Vink} J.~S.,  {Eldridge} J.~J.,   {Tout} C.~A.,  2019, \mn@doi [\mnras] {10.1093/mnras/stz1134}, \href {https://ui.adsabs.harvard.edu/abs/2019MNRAS.486.4451G} {486, 4451}

\bibitem[\protect\citeauthoryear{{Goswami} et~al.,}{{Goswami} et~al.}{2021}]{Goswami}
{Goswami} S.,  et~al., 2021, \mn@doi [\aap] {10.1051/0004-6361/202039842}, \href {https://ui.adsabs.harvard.edu/abs/2021A&A...650A.203G} {650, A203}

\bibitem[\protect\citeauthoryear{{G{\"o}tberg}, {Korol}, {Lamberts}, {Kupfer}, {Breivik}, {Ludwig}  \& {Drout}}{{G{\"o}tberg} et~al.}{2020}]{goetberg20}
{G{\"o}tberg} Y.,  {Korol} V.,  {Lamberts} A.,  {Kupfer} T.,  {Breivik} K.,  {Ludwig} B.,   {Drout} M.~R.,  2020, \mn@doi [\apj] {10.3847/1538-4357/abbda5}, \href {https://ui.adsabs.harvard.edu/abs/2020ApJ...904...56G} {904, 56}

\bibitem[\protect\citeauthoryear{{Gr{\"a}fener}, {Vink}, {Harries}  \& {Langer}}{{Gr{\"a}fener} et~al.}{2012}]{GVHL12}
{Gr{\"a}fener} G.,  {Vink} J.~S.,  {Harries} T.~J.,   {Langer} N.,  2012, \mn@doi [\aap] {10.1051/0004-6361/201118664}, \href {https://ui.adsabs.harvard.edu/abs/2012A&A...547A..83G} {547, A83}

\bibitem[\protect\citeauthoryear{{Groh} et~al.,}{{Groh} et~al.}{2019}]{groh19}
{Groh} J.~H.,  et~al., 2019, \mn@doi [\aap] {10.1051/0004-6361/201833720}, \href {https://ui.adsabs.harvard.edu/abs/2019A&A...627A..24G} {627, A24}

\bibitem[\protect\citeauthoryear{{Hainich} et~al.,}{{Hainich} et~al.}{2014}]{Hainich14}
{Hainich} R.,  et~al., 2014, \mn@doi [\aap] {10.1051/0004-6361/201322696}, \href {https://ui.adsabs.harvard.edu/abs/2014A&A...565A..27H} {565, A27}

\bibitem[\protect\citeauthoryear{{Hainich}, {Pasemann}, {Todt}, {Shenar}, {Sander}  \& {Hamann}}{{Hainich} et~al.}{2015}]{Hainich15}
{Hainich} R.,  {Pasemann} D.,  {Todt} H.,  {Shenar} T.,  {Sander} A.,   {Hamann} W.~R.,  2015, \mn@doi [\aap] {10.1051/0004-6361/201526241}, \href {https://ui.adsabs.harvard.edu/abs/2015A&A...581A..21H} {581, A21}

\bibitem[\protect\citeauthoryear{{Hamann}, {Gr{\"a}fener}  \& {Liermann}}{{Hamann} et~al.}{2006}]{Hamann06}
{Hamann} W.~R.,  {Gr{\"a}fener} G.,   {Liermann} A.,  2006, \mn@doi [\aap] {10.1051/0004-6361:20065052}, \href {https://ui.adsabs.harvard.edu/abs/2006A&A...457.1015H} {457, 1015}

\bibitem[\protect\citeauthoryear{{Hamann} et~al.,}{{Hamann} et~al.}{2019}]{Hamann+2019}
{Hamann} W.~R.,  et~al., 2019, \mn@doi [\aap] {10.1051/0004-6361/201834850}, \href {https://ui.adsabs.harvard.edu/abs/2019A&A...625A..57H} {625, A57}

\bibitem[\protect\citeauthoryear{{Heger}, {Langer}  \& {Woosley}}{{Heger} et~al.}{2000}]{Heger00}
{Heger} A.,  {Langer} N.,   {Woosley} S.~E.,  2000, \mn@doi [\apj] {10.1086/308158}, \href {https://ui.adsabs.harvard.edu/abs/2000ApJ...528..368H} {528, 368}

\bibitem[\protect\citeauthoryear{{Herwig}}{{Herwig}}{2000}]{Herwig2000}
{Herwig} F.,  2000, \aap, 360, 952

\bibitem[\protect\citeauthoryear{{Higdon} \& {Lingenfelter}}{{Higdon} \& {Lingenfelter}}{2003}]{HigLing03}
{Higdon} J.~C.,  {Lingenfelter} R.~E.,  2003, \mn@doi [\apj] {10.1086/375192}, \href {https://ui.adsabs.harvard.edu/abs/2003ApJ...590..822H} {590, 822}

\bibitem[\protect\citeauthoryear{{Higgins}, {Sander}, {Vink}  \& {Hirschi}}{{Higgins} et~al.}{2021}]{Higgins+21}
{Higgins} E.~R.,  {Sander} A.~A.~C.,  {Vink} J.~S.,   {Hirschi} R.,  2021, \mn@doi [\mnras] {10.1093/mnras/stab1548}, \href {https://ui.adsabs.harvard.edu/abs/2021MNRAS.505.4874H} {505, 4874}

\bibitem[\protect\citeauthoryear{{Higgins}, {Vink}, {Sabhahit}  \& {Sander}}{{Higgins} et~al.}{2022}]{Higgins22}
{Higgins} E.~R.,  {Vink} J.~S.,  {Sabhahit} G.~N.,   {Sander} A. A.~C.,  2022, \mn@doi [\mnras] {10.1093/mnras/stac2485}, \href {https://ui.adsabs.harvard.edu/abs/2022MNRAS.516.4052H} {516, 4052}

\bibitem[\protect\citeauthoryear{{Higgins}, {Vink}, {Hirschi}, {Laird}  \& {Sabhahit}}{{Higgins} et~al.}{2023}]{Higgins+23}
{Higgins} E.~R.,  {Vink} J.~S.,  {Hirschi} R.,  {Laird} A.~M.,   {Sabhahit} G.~N.,  2023, \mn@doi [\mnras] {10.1093/mnras/stad2537}, \href {https://ui.adsabs.harvard.edu/abs/2023MNRAS.526..534H} {526, 534}

\bibitem[\protect\citeauthoryear{{Hirschi}, {Meynet}  \& {Maeder}}{{Hirschi} et~al.}{2005}]{hirschi05}
{Hirschi} R.,  {Meynet} G.,   {Maeder} A.,  2005, \mn@doi [\aap] {10.1051/0004-6361:20041554}, \href {https://ui.adsabs.harvard.edu/abs/2005A&A...433.1013H} {433, 1013}

\bibitem[\protect\citeauthoryear{{Iliadis}}{{Iliadis}}{2010}]{iliadis}
{Iliadis} C.,  2010, in {Spitaleri} C.,  {Rolfs} C.,   {Pizzone} R.~G.,  eds,  American Institute of Physics Conference Series Vol. 1213, Fifth European Summer School on Experimental Nuclear AstroPhysics. pp 3--22 (\mn@eprint {arXiv} {0911.3965}), \mn@doi{10.1063/1.3362604}

\bibitem[\protect\citeauthoryear{{Josiek}, {Ekstr{\"o}m}  \& {Sander}}{{Josiek} et~al.}{2024}]{joris24}
{Josiek} J.,  {Ekstr{\"o}m} S.,   {Sander} A. A.~C.,  2024, arXiv e-prints, \href {https://ui.adsabs.harvard.edu/abs/2024arXiv240414488J} {p. arXiv:2404.14488}

\bibitem[\protect\citeauthoryear{{Klencki}, {Nelemans}, {Istrate}  \& {Pols}}{{Klencki} et~al.}{2020}]{Klencki+2020}
{Klencki} J.,  {Nelemans} G.,  {Istrate} A.~G.,   {Pols} O.,  2020, \mn@doi [\aap] {10.1051/0004-6361/202037694}, \href {https://ui.adsabs.harvard.edu/abs/2020A&A...638A..55K} {638, A55}

\bibitem[\protect\citeauthoryear{{Langer}}{{Langer}}{1989}]{langer89}
{Langer} N.,  1989, \aap, \href {https://ui.adsabs.harvard.edu/abs/1989A&A...220..135L} {220, 135}

\bibitem[\protect\citeauthoryear{{Laplace}, {G{\"o}tberg}, {de Mink}, {Justham}  \& {Farmer}}{{Laplace} et~al.}{2020}]{Laplace+2020}
{Laplace} E.,  {G{\"o}tberg} Y.,  {de Mink} S.~E.,  {Justham} S.,   {Farmer} R.,  2020, \mn@doi [\aap] {10.1051/0004-6361/201937300}, \href {https://ui.adsabs.harvard.edu/abs/2020A&A...637A...6L} {637, A6}

\bibitem[\protect\citeauthoryear{{Limongi} \& {Chieffi}}{{Limongi} \& {Chieffi}}{2006}]{limongichieffi06}
{Limongi} M.,  {Chieffi} A.,  2006, \mn@doi [\apj] {10.1086/505164}, \href {https://ui.adsabs.harvard.edu/abs/2006ApJ...647..483L} {647, 483}

\bibitem[\protect\citeauthoryear{{Lingenfelter}, {Higdon}  \& {Ramaty}}{{Lingenfelter} et~al.}{2000}]{linghig00}
{Lingenfelter} R.~E.,  {Higdon} J.~C.,   {Ramaty} R.,  2000, in {Mewaldt} R.~A.,  {Jokipii} J.~R.,  {Lee} M.~A.,  {M{\"o}bius} E.,   {Zurbuchen} T.~H.,  eds,  American Institute of Physics Conference Series Vol. 528, Acceleration and Transport of Energetic Particles Observed in the Heliosphere. pp 375--382 (\mn@eprint {arXiv} {astro-ph/0004166}), \mn@doi{10.1063/1.1324342}

\bibitem[\protect\citeauthoryear{{Lukasiak}, {Ferrando}, {McDonald}  \& {Webber}}{{Lukasiak} et~al.}{1994}]{lukasiak94}
{Lukasiak} A.,  {Ferrando} P.,  {McDonald} F.~B.,   {Webber} W.~R.,  1994, \mn@doi [\apj] {10.1086/174072}, \href {https://ui.adsabs.harvard.edu/abs/1994ApJ...426..366L} {426, 366}

\bibitem[\protect\citeauthoryear{{Maeder}}{{Maeder}}{1992}]{maeder92}
{Maeder} A.,  1992, \aap, \href {https://ui.adsabs.harvard.edu/abs/1992A&A...264..105M} {264, 105}

\bibitem[\protect\citeauthoryear{{Maeder} \& {Meynet}}{{Maeder} \& {Meynet}}{2012}]{mm12}
{Maeder} A.,  {Meynet} G.,  2012, \mn@doi [Reviews of Modern Physics] {10.1103/RevModPhys.84.25}, \href {https://ui.adsabs.harvard.edu/abs/2012RvMP...84...25M} {84, 25}

\bibitem[\protect\citeauthoryear{{Martinet} et~al.,}{{Martinet} et~al.}{2022}]{martinet22}
{Martinet} S.,  et~al., 2022, \mn@doi [\aap] {10.1051/0004-6361/202243474}, \href {https://ui.adsabs.harvard.edu/abs/2022A&A...664A.181M} {664, A181}

\bibitem[\protect\citeauthoryear{{Martins}}{{Martins}}{2015}]{martins15}
{Martins} F.,  2015, in {Vink} J.~S.,  ed.,  Astrophysics and Space Science Library Vol. 412, Very Massive Stars in the Local Universe. p.~9 (\mn@eprint {arXiv} {1404.0166}), \mn@doi{10.1007/978-3-319-09596-7\_2}

\bibitem[\protect\citeauthoryear{{McClelland} \& {Eldridge}}{{McClelland} \& {Eldridge}}{2016}]{mcclellannd16}
{McClelland} L.~A.~S.,  {Eldridge} J.~J.,  2016, \mn@doi [\mnras] {10.1093/mnras/stw618}, \href {https://ui.adsabs.harvard.edu/abs/2016MNRAS.459.1505M} {459, 1505}

\bibitem[\protect\citeauthoryear{{Meynet}}{{Meynet}}{2008}]{meynet08}
{Meynet} G.,  2008, \mn@doi [European Physical Journal Special Topics] {10.1140/epjst/e2008-00623-1}, \href {https://ui.adsabs.harvard.edu/abs/2008EPJST.156..257M} {156, 257}

\bibitem[\protect\citeauthoryear{{Meynet} \& {Arnould}}{{Meynet} \& {Arnould}}{2000}]{meynetarnould00}
{Meynet} G.,  {Arnould} M.,  2000, \mn@doi [\aap] {10.48550/arXiv.astro-ph/0001170}, \href {https://ui.adsabs.harvard.edu/abs/2000A&A...355..176M} {355, 176}

\bibitem[\protect\citeauthoryear{{Neugent} \& {Massey}}{{Neugent} \& {Massey}}{2019}]{NeugentMassey19}
{Neugent} K.,  {Massey} P.,  2019, \mn@doi [Galaxies] {10.3390/galaxies7030074}, \href {https://ui.adsabs.harvard.edu/abs/2019Galax...7...74N} {7, 74}

\bibitem[\protect\citeauthoryear{{Nugis} \& {Lamers}}{{Nugis} \& {Lamers}}{2000}]{NugisLamers}
{Nugis} T.,  {Lamers} H.~J.~G.~L.~M.,  2000, \aap, \href {https://ui.adsabs.harvard.edu/abs/2000A&A...360..227N} {360, 227}

\bibitem[\protect\citeauthoryear{{O'Connor} \& {Ott}}{{O'Connor} \& {Ott}}{2011}]{oconnor}
{O'Connor} E.,  {Ott} C.~D.,  2011, \mn@doi [\apj] {10.1088/0004-637X/730/2/70}, \href {http://ukads.nottingham.ac.uk/abs/2011ApJ...730...70O} {730, 70}

\bibitem[\protect\citeauthoryear{{Olive} \& {Vangioni}}{{Olive} \& {Vangioni}}{2019}]{olive19}
{Olive} K.~A.,  {Vangioni} E.,  2019, \mn@doi [\mnras] {10.1093/mnras/stz2893}, \href {https://ui.adsabs.harvard.edu/abs/2019MNRAS.490.4307O} {490, 4307}

\bibitem[\protect\citeauthoryear{{Paczy{\'n}ski}}{{Paczy{\'n}ski}}{1967}]{pacz67}
{Paczy{\'n}ski} B.,  1967, \actaa, \href {https://ui.adsabs.harvard.edu/abs/1967AcA....17..355P} {17, 355}

\bibitem[\protect\citeauthoryear{{Palacios}, {Arnould}  \& {Meynet}}{{Palacios} et~al.}{2005}]{palacios05}
{Palacios} A.,  {Arnould} M.,   {Meynet} G.,  2005, \mn@doi [\aap] {10.1051/0004-6361:20053323}, \href {https://ui.adsabs.harvard.edu/abs/2005A&A...443..243P} {443, 243}

\bibitem[\protect\citeauthoryear{{Paxton}, {Bildsten}, {Dotter}, {Herwig}, {Lesaffre}  \& {Timmes}}{{Paxton} et~al.}{2011}]{Pax11}
{Paxton} B.,  {Bildsten} L.,  {Dotter} A.,  {Herwig} F.,  {Lesaffre} P.,   {Timmes} F.,  2011, \apjs, 192, 3

\bibitem[\protect\citeauthoryear{{Paxton} et~al.,}{{Paxton} et~al.}{2013}]{Pax13}
{Paxton} B.,  et~al., 2013, \apjs, 208, 4

\bibitem[\protect\citeauthoryear{{Paxton} et~al.,}{{Paxton} et~al.}{2015}]{Pax15}
{Paxton} B.,  et~al., 2015, \apjs, 220, 15

\bibitem[\protect\citeauthoryear{{Paxton} et~al.,}{{Paxton} et~al.}{2018}]{Pax18}
{Paxton} B.,  et~al., 2018, \mn@doi [\apjs] {10.3847/1538-4365/aaa5a8}, \href {https://ui.adsabs.harvard.edu/abs/2018ApJS..234...34P} {234, 34}

\bibitem[\protect\citeauthoryear{{Paxton} et~al.,}{{Paxton} et~al.}{2019}]{Pax19}
{Paxton} B.,  et~al., 2019, \mn@doi [\apjs] {10.3847/1538-4365/ab2241}, \href {https://ui.adsabs.harvard.edu/abs/2019ApJS..243...10P} {243, 10}

\bibitem[\protect\citeauthoryear{{Podsiadlowski}, {Joss}  \& {Hsu}}{{Podsiadlowski} et~al.}{1992}]{Podsi92}
{Podsiadlowski} P.,  {Joss} P.~C.,   {Hsu} J.~J.~L.,  1992, \mn@doi [\apj] {10.1086/171341}, \href {https://ui.adsabs.harvard.edu/abs/1992ApJ...391..246P} {391, 246}

\bibitem[\protect\citeauthoryear{{Pols} \& {Dewi}}{{Pols} \& {Dewi}}{2002}]{pols02}
{Pols} O.~R.,  {Dewi} J. D.~M.,  2002, \mn@doi [\pasa] {10.1071/AS01121}, \href {https://ui.adsabs.harvard.edu/abs/2002PASA...19..233P} {19, 233}

\bibitem[\protect\citeauthoryear{{Renda} et~al.,}{{Renda} et~al.}{2004}]{renda04}
{Renda} A.,  et~al., 2004, \mn@doi [\mnras] {10.1111/j.1365-2966.2004.08215.x}, \href {https://ui.adsabs.harvard.edu/abs/2004MNRAS.354..575R} {354, 575}

\bibitem[\protect\citeauthoryear{Rogers \& Nayfonov}{Rogers \& Nayfonov}{2002}]{RogersNayfonov02}
Rogers F.,  Nayfonov A.,  2002, The Astrophysical Journal, 576, 1064

\bibitem[\protect\citeauthoryear{Rosslowe \& Crowther}{Rosslowe \& Crowther}{2015}]{rosslowecrow05}
Rosslowe C.~K.,  Crowther P.~A.,  2015, \mn@doi [Monthly Notices of the Royal Astronomical Society] {10.1093/mnras/stu2525}, 447, 2322

\bibitem[\protect\citeauthoryear{{Ryde} et~al.,}{{Ryde} et~al.}{2020}]{Ryde20}
{Ryde} N.,  et~al., 2020, \mn@doi [\apj] {10.3847/1538-4357/ab7eb1}, \href {https://ui.adsabs.harvard.edu/abs/2020ApJ...893...37R} {893, 37}

\bibitem[\protect\citeauthoryear{{Sabhahit}, {Vink}, {Higgins}  \& {Sander}}{{Sabhahit} et~al.}{2022}]{Sabh22}
{Sabhahit} G.~N.,  {Vink} J.~S.,  {Higgins} E.~R.,   {Sander} A. A.~C.,  2022, \mn@doi [\mnras] {10.1093/mnras/stac1410}, \href {https://ui.adsabs.harvard.edu/abs/2022MNRAS.514.3736S} {514, 3736}

\bibitem[\protect\citeauthoryear{{Sander} \& {Vink}}{{Sander} \& {Vink}}{2020}]{SV2020}
{Sander} A. A.~C.,  {Vink} J.~S.,  2020, \mn@doi [\mnras] {10.1093/mnras/staa2712}, \href {https://ui.adsabs.harvard.edu/abs/2020MNRAS.499..873S} {499, 873}

\bibitem[\protect\citeauthoryear{{Sander}, {Hamann}  \& {Todt}}{{Sander} et~al.}{2012}]{Sander12}
{Sander} A.,  {Hamann} W.~R.,   {Todt} H.,  2012, \mn@doi [\aap] {10.1051/0004-6361/201117830}, \href {https://ui.adsabs.harvard.edu/abs/2012A&A...540A.144S} {540, A144}

\bibitem[\protect\citeauthoryear{Sander, Hamann, Todt, Hainich, Shenar, Ramachandran  \& Oskinova}{Sander et~al.}{2019}]{Sander19}
Sander A.~A.,  Hamann W.-R.,  Todt H.,  Hainich R.,  Shenar T.,  Ramachandran V.,   Oskinova L.~M.,  2019, Astronomy \& Astrophysics, 621, A92

\bibitem[\protect\citeauthoryear{{Sander}, {Lefever}, {Poniatowski}, {Ramachandran}, {Sabhahit}  \& {Vink}}{{Sander} et~al.}{2023}]{Sander23}
{Sander} A.~A.~C.,  {Lefever} R.~R.,  {Poniatowski} L.~G.,  {Ramachandran} V.,  {Sabhahit} G.~N.,   {Vink} J.~S.,  2023, \mn@doi [\aap] {10.1051/0004-6361/202245110}, \href {https://ui.adsabs.harvard.edu/abs/2023A&A...670A..83S} {670, A83}

\bibitem[\protect\citeauthoryear{{Spitoni}, {Matteucci}, {J{\"o}nsson}, {Ryde}  \& {Romano}}{{Spitoni} et~al.}{2018}]{spitoni}
{Spitoni} E.,  {Matteucci} F.,  {J{\"o}nsson} H.,  {Ryde} N.,   {Romano} D.,  2018, \mn@doi [\aap] {10.1051/0004-6361/201732092}, \href {https://ui.adsabs.harvard.edu/abs/2018A&A...612A..16S} {612, A16}

\bibitem[\protect\citeauthoryear{{Tatischeff}, {Duprat}  \& {de S{\'e}r{\'e}ville}}{{Tatischeff} et~al.}{2010}]{tatischeff10}
{Tatischeff} V.,  {Duprat} J.,   {de S{\'e}r{\'e}ville} N.,  2010, \mn@doi [\apjl] {10.1088/2041-8205/714/1/L26}, \href {https://ui.adsabs.harvard.edu/abs/2010ApJ...714L..26T} {714, L26}

\bibitem[\protect\citeauthoryear{{Thielemann} \& {Arnett}}{{Thielemann} \& {Arnett}}{1985}]{thiele85}
{Thielemann} F.~K.,  {Arnett} W.~D.,  1985, \mn@doi [\apj] {10.1086/163403}, \href {https://ui.adsabs.harvard.edu/abs/1985ApJ...295..604T} {295, 604}

\bibitem[\protect\citeauthoryear{{Tramper} et~al.,}{{Tramper} et~al.}{2015}]{Tramper15}
{Tramper} F.,  et~al., 2015, \mn@doi [\aap] {10.1051/0004-6361/201425390}, \href {https://ui.adsabs.harvard.edu/abs/2015A&A...581A.110T} {581, A110}

\bibitem[\protect\citeauthoryear{{Vink} \& {Gr{\"a}fener}}{{Vink} \& {Gr{\"a}fener}}{2012}]{VG12}
{Vink} J.~S.,  {Gr{\"a}fener} G.,  2012, \mn@doi [\apjl] {10.1088/2041-8205/751/2/L34}, \href {http://ukads.nottingham.ac.uk/abs/2012ApJ...751L..34V} {751, L34}

\bibitem[\protect\citeauthoryear{{Vink} \& {Harries}}{{Vink} \& {Harries}}{2017}]{VinkHarries}
{Vink} J.~S.,  {Harries} T.~J.,  2017, \mn@doi [\aap] {10.1051/0004-6361/201730503}, \href {http://ukads.nottingham.ac.uk/abs/2017A%26A...603A.120V} {603, A120}

\bibitem[\protect\citeauthoryear{Vink \& de Koter}{Vink \& de~Koter}{2005}]{Vink05}
Vink J.~S.,  de Koter A.,  2005, Astronomy \& Astrophysics, 442, 587

\bibitem[\protect\citeauthoryear{{Vink}, {Muijres}, {Anthonisse}, {de Koter}, {Gr{\"a}fener}  \& {Langer}}{{Vink} et~al.}{2011a}]{vink11}
{Vink} J.~S.,  {Muijres} L.~E.,  {Anthonisse} B.,  {de Koter} A.,  {Gr{\"a}fener} G.,   {Langer} N.,  2011a, \mn@doi [\aap] {10.1051/0004-6361/201116614}, \href {https://ui.adsabs.harvard.edu/abs/2011A&A...531A.132V} {531, A132}

\bibitem[\protect\citeauthoryear{{Vink}, {Gr{\"a}fener}  \& {Harries}}{{Vink} et~al.}{2011b}]{vink11b}
{Vink} J.~S.,  {Gr{\"a}fener} G.,   {Harries} T.~J.,  2011b, \mn@doi [\aap] {10.1051/0004-6361/201118197}, \href {https://ui.adsabs.harvard.edu/abs/2011A&A...536L..10V} {536, L10}

\bibitem[\protect\citeauthoryear{{Vink} et~al.,}{{Vink} et~al.}{2015}]{vinkbook}
{Vink} J.~S.,  et~al., 2015, Highlights of Astronomy, 16, 51

\bibitem[\protect\citeauthoryear{{Wiedenbeck} \& {Greiner}}{{Wiedenbeck} \& {Greiner}}{1981}]{weidenbeck81}
{Wiedenbeck} M.~E.,  {Greiner} D.~E.,  1981, \mn@doi [\prl] {10.1103/PhysRevLett.46.682}, \href {https://ui.adsabs.harvard.edu/abs/1981PhRvL..46..682W} {46, 682}

\bibitem[\protect\citeauthoryear{{Wolf} \& {Rayet}}{{Wolf} \& {Rayet}}{1867}]{WR67}
{Wolf} C.~J.~E.,  {Rayet} G.,  1867, Academie des Sciences Paris Comptes Rendus, \href {https://ui.adsabs.harvard.edu/abs/1867CRAS...65..292W} {65, 292}

\bibitem[\protect\citeauthoryear{{Woosley}}{{Woosley}}{2019}]{Woosley19}
{Woosley} S.~E.,  2019, \mn@doi [\apj] {10.3847/1538-4357/ab1b41}, \href {https://ui.adsabs.harvard.edu/abs/2019ApJ...878...49W} {878, 49}

\bibitem[\protect\citeauthoryear{{Woosley} \& {Heger}}{{Woosley} \& {Heger}}{2006}]{woosleyHeger06}
{Woosley} S.~E.,  {Heger} A.,  2006, \mn@doi [\apj] {10.1086/498500}, \href {https://ui.adsabs.harvard.edu/abs/2006ApJ...637..914W} {637, 914}

\bibitem[\protect\citeauthoryear{{Yoon} \& {Langer}}{{Yoon} \& {Langer}}{2005}]{Yoon05}
{Yoon} S.~C.,  {Langer} N.,  2005, \mn@doi [\aap] {10.1051/0004-6361:20054030}, \href {https://ui.adsabs.harvard.edu/abs/2005A&A...443..643Y} {443, 643}

\bibitem[\protect\citeauthoryear{{de Jager}, {Nieuwenhuijzen}  \& {van der Hucht}}{{de Jager} et~al.}{1988}]{dejager88}
{de Jager} C.,  {Nieuwenhuijzen} H.,   {van der Hucht} K.~A.,  1988, \aaps, \href {https://ui.adsabs.harvard.edu/abs/1988A&AS...72..259D} {72, 259}

\bibitem[\protect\citeauthoryear{{van der Hucht}}{{van der Hucht}}{2001}]{vanderHucht01}
{van der Hucht} K.~A.,  2001, \mn@doi [\nar] {10.1016/S1387-6473(00)00112-3}, \href {https://ui.adsabs.harvard.edu/abs/2001NewAR..45..135V} {45, 135}

\makeatother
\end{thebibliography}
\begin{appendix}
\section{Ejected masses of 22 isotopes}\label{92yieldapp}
\begin{table}
    \centering
    \begin{tabular}{c|c|c|c}
    \hline
          Isotope &  Ejected mass & Isotope &  Ejected mass\\
        \hline \hline
H$^{1}$	    &1.27E-04	 &Ne$^{20}$	&1.71E-02				\\
He$^{3}$	&1.15E-16 &Ne$^{21}$	&2.39E-05			\\
He$^{4}$	&8.78E+00	    &Ne$^{22}$	&7.85E-02			\\
C$^{12}$	&1.88E+00&Na$^{23}$	&3.24E-03		\\
C$^{13}$	&2.14E-04&Mg$^{24}$	&5.54E-03			\\
N$^{14}$	&4.20E-02	 &Mg$^{25}$	&3.87E-04			\\
N$^{15}	$   &1.69E-06	   &Mg$^{26}$	&1.68E-03	\\
O$^{16}$	&4.01E-01 &Al$^{26}$	&1.17E-04\\
O$^{17}	$&2.18E-06	 &Al$^{27}$	&8.22E-04		\\
O$^{18}	$&9.46E-05 &Si$^{28}$	&6.53E-03				\\
F$^{19}$	&4.18E-05			 &Si$^{30}$	&3.19E-04		\\

      \hline
    \end{tabular}
    \caption{Ejected masses for a 30\Mdot\ classical WR model, calculated from the onset of core He-burning until core O-exhaustion. }
    \label{tab:92yield30}    
\end{table}

\section {Figures}
%RH
Additional figures are presented in this Appendix.

\begin{figure*}
    \centering
    \includegraphics[width = \textwidth]{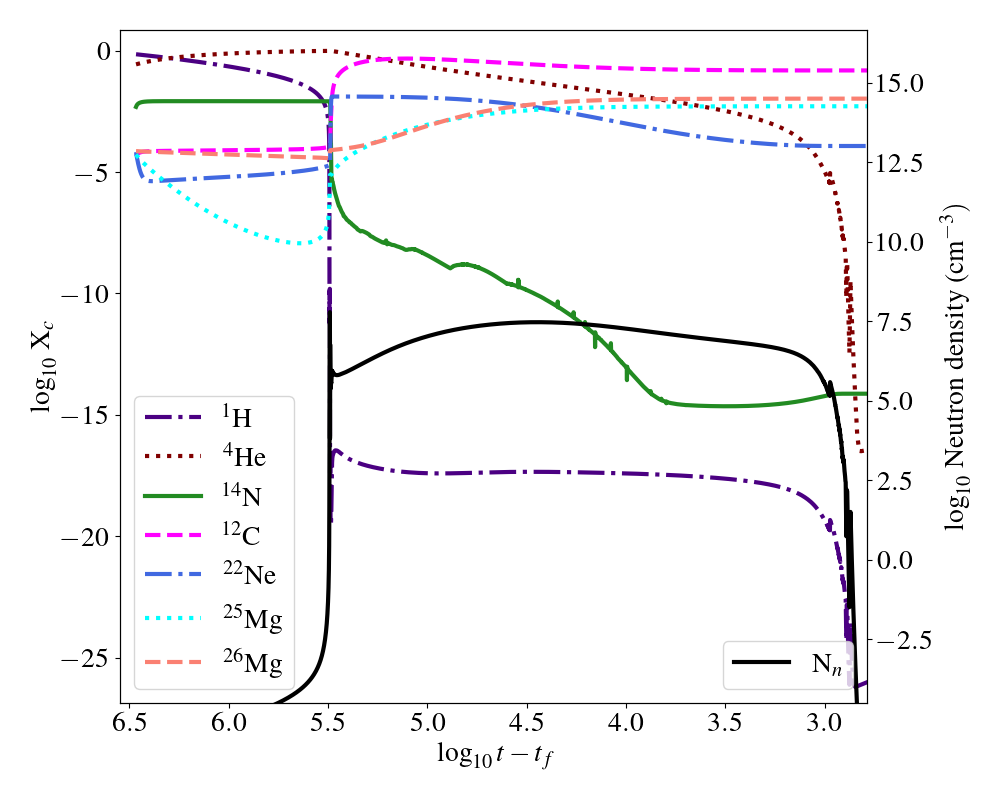}
    \caption{Evolution of the central composition (left axis) and neutron density (right axis), with time in log-scale from core H-burning until core He-exhaustion for a 200\Mdot\ star. }
    \label{fig:200_neut_Xc}
\end{figure*}

\begin{figure}
    \centering
    \includegraphics[width = \columnwidth]{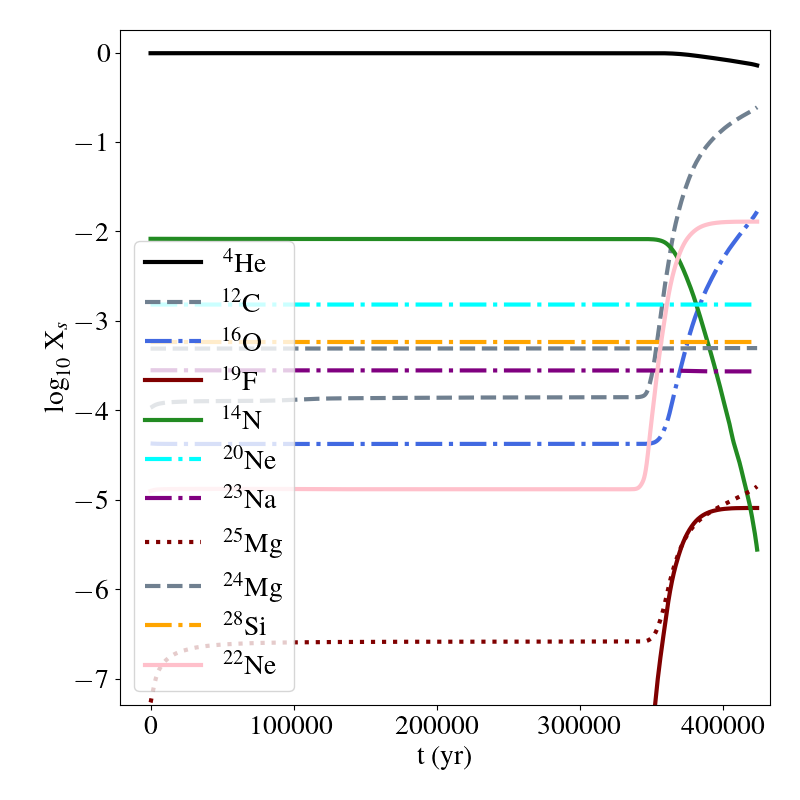}
    \caption{Time evolution of the surface composition during core He-, C- and O-burning phases, for a model with an initial mass of 20\Mdot. }
    \label{fig:20_Xs}
\end{figure}
\begin{figure}
    \centering
    \includegraphics[width = \columnwidth]{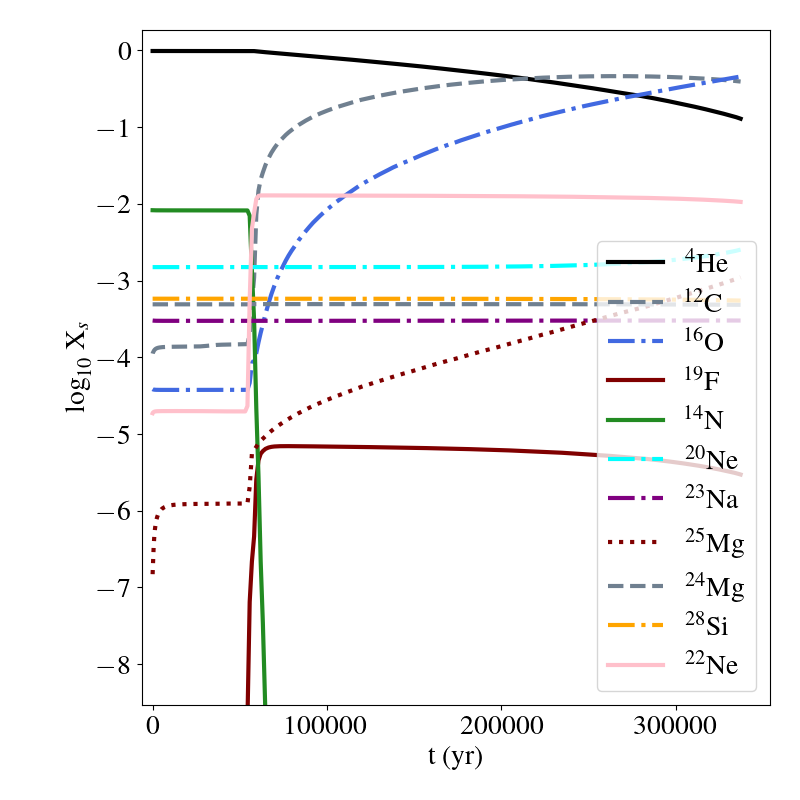}
    \caption{Time evolution of the surface composition during core He-, C- and O-burning phases, for a model with an initial mass of 50\Mdot. }
    \label{fig:50_Xs}
\end{figure}

\begin{figure}
    \centering
    \includegraphics[width = \columnwidth]{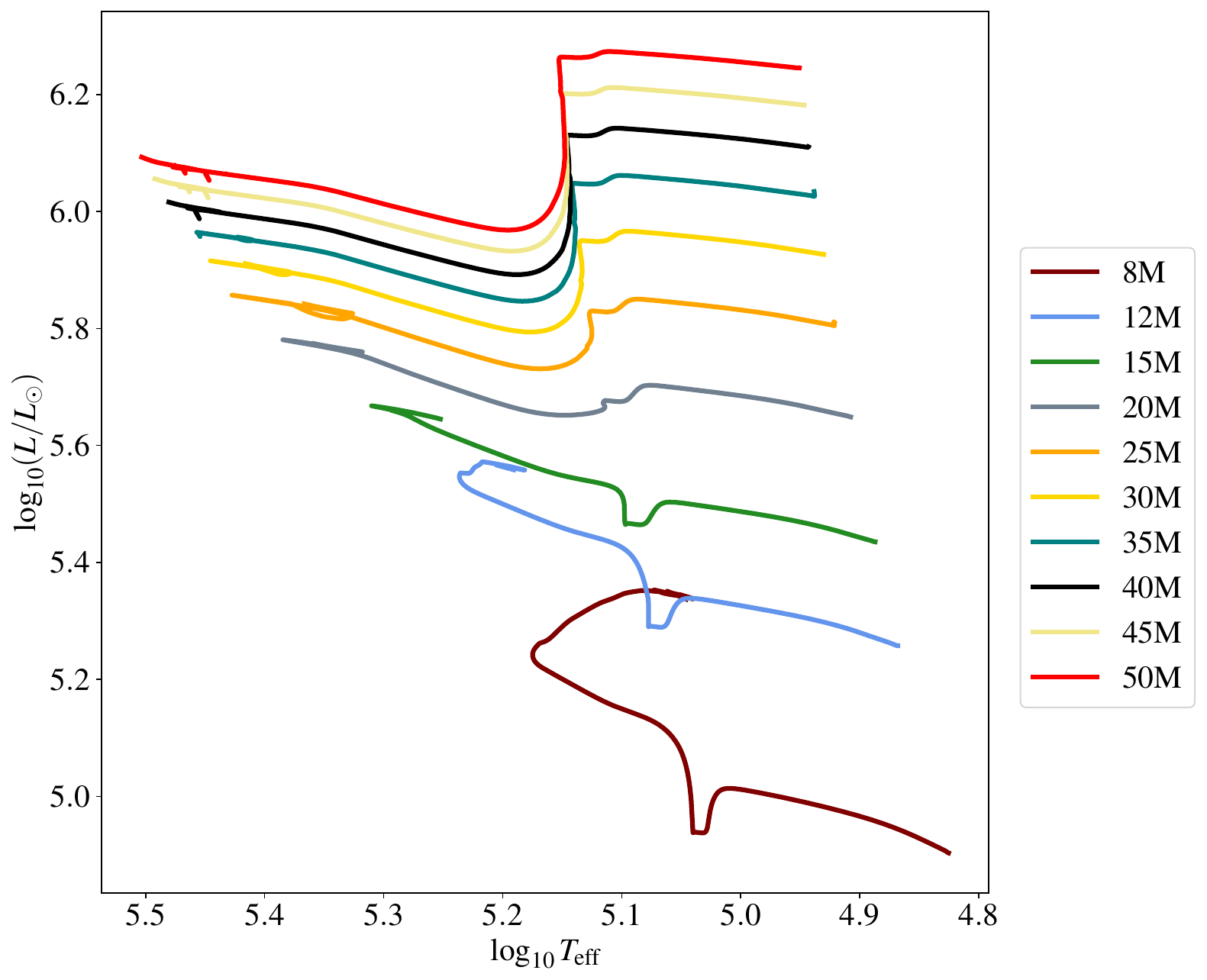}
    \caption{Hertzsprung-Russell diagram of our grid of models for a range of initial masses, calculated from core He-burning until core O-exhaustion.}
    \label{fig:HRD}
\end{figure}
\begin{figure}
    \centering
    \includegraphics[width = \columnwidth]{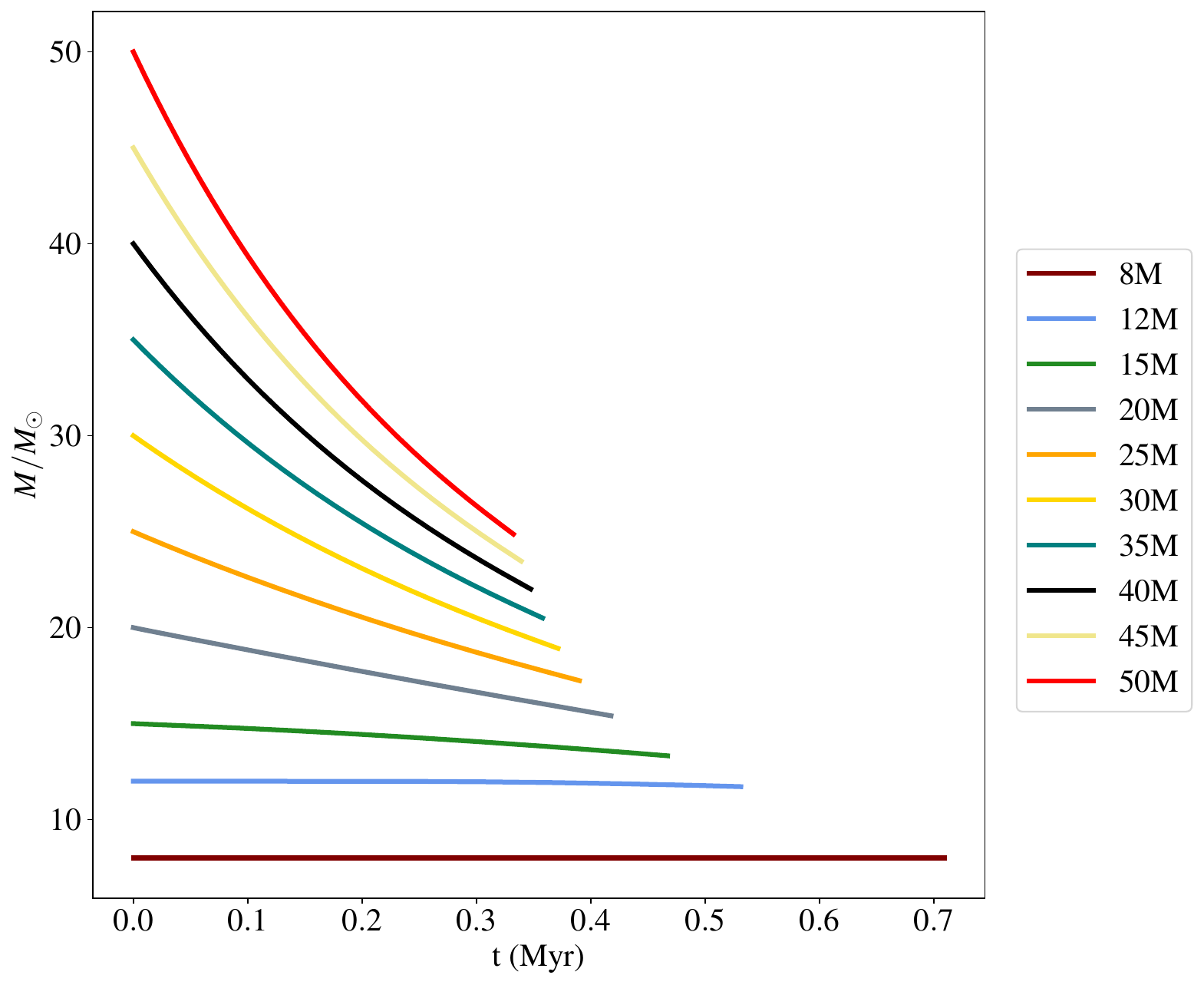}
    \caption{Mass evolution of our grid of models, shown for the complete evolution.}
    \label{fig:Mt}
\end{figure}

\end{appendix}
\end{document}